\documentclass[useAMS,usenatbib]{mn2e}

\usepackage{epsfig}
\usepackage{amsmath}
\usepackage{amssymb}
\usepackage{natbib}
\usepackage{threeparttable} 
\usepackage{booktabs}
\usepackage{float}
\usepackage{times}

\usepackage{epstopdf}

\newcommand{\chan}{\textit{Chandra}}
\newcommand{\swift}{\textit{Swift}}

\newcommand{\xmm}{\textit{XMM-Newton}}

\newcommand{\maxi}{\textit{MAXI}}

\newcommand{\Msun}{\mathrm{M}_{\odot}}
\newcommand{\lum}{\mathrm{erg~s}^{-1}}
\newcommand{\flux}{\mathrm{erg~cm}^{-2}~\mathrm{s}^{-1}}
\newcommand{\fluence}{\mathrm{erg~cm}^{-2}}
\newcommand{\cnts}{\mathrm{c~s}^{-1}}

\newcommand{\mdotg}{\mathrm{g~s}^{-1}}
\newcommand{\nh}{\mathrm{cm}^{-2}}
\newcommand{\dist}{(D/5.0~\mathrm{kpc})^2}

\newcommand{\source}{Aql X-1}
\newcommand{\xte}{XTE J1701--462}
\newcommand{\exo}{EXO~0748--676}
\newcommand{\ks}{KS~1731--260}
\newcommand{\mxb}{MXB~1659--29}

\newcommand{\maxisource}{MAXI~J0556--332}
\newcommand{\terdrie}{Swift~J174805.3--244637}
\newcommand{\igr}{IGR J17480--2446}

\newcommand{\sax}{SAX J1808.4--3658}
\newcommand{\exoter}{EXO 1745--248}

\def \mnras {MNRAS}
\def \apj {ApJ}
\def \apjs {ApJS}
\def \apjl {ApJL}
\def \aap {A\&A}
\def \nat {Nature}

\def \pasj {PASJ}

\def \prc {PhRvC}

\def \pre {PhRvE}

\title[Crust Cooling in \source]{Constraining the properties of neutron star crusts with the transient low-mass X-ray binary \source}
\author[A.C. Waterhouse et al.]
{A.~C. Waterhouse$^{1}$, N.~Degenaar$^1$\thanks{e-mail: degenaar@ast.cam.ac.uk}, R. Wijnands$^{2}$, E.~F.~Brown$^{3}$, J.~M.~Miller$^{4}$, 
\newauthor D.~Altamirano$^{5}$, and M.~Linares$^{6,7,8}$\\
$^1$Institute of Astronomy, University of Cambridge, Madingley Road, Cambridge CB3 OHA, UK\\
$^2$Anton Pannekoek Institute for Astronomy, University of Amsterdam, Postbus 94249, 1090 GE Amsterdam, The Netherlands\\
$^3$Department of Physics and Astronomy, Michigan State University, East Lansing, MI 48824, USA\\
$^4$Department of Astronomy, University of Michigan, 500 Church Street, Ann Arbor, MI 48109, USA\\
$^5$Department of Physics and Astronomy, Southampton University, Southampton SO17 1BJ, UK\\
$^6$Instituto de Astrof\'{i}sica de Canarias, c/ V\'{i}a L\'{a}ctea s/n, E-38205 La Laguna, Tenerife, Spain\\
$^7$Universidad de La Laguna, Departamento de Astrof\'{i}sica, E-38206 La Laguna, Tenerife, Spain\\
$^8$Institutt for fysikk, NTNU, Trondheim, Norway
}

\begin{document}

\date{Accepted 2015 December 16. Received 2015 December 15; in original form 2015 October 5}

\pagerange{\pageref{firstpage}--\pageref{lastpage}} \pubyear{0000}

\maketitle

\label{firstpage}

\begin{abstract}
\source\ is a prolific transient neutron star low-mass X-ray binary that exhibits an accretion outburst approximately once every year.
Whether the thermal X-rays detected in intervening quiescent episodes are the result of cooling of the neutron star or due to continued low-level accretion remains unclear. In this work we use \swift\ data obtained after the long and bright 2011 and 2013 outbursts, as well as the short and faint 2015 outburst, to investigate the hypothesis that cooling of the accretion-heated neutron star crust dominates the quiescent thermal emission in \source. We demonstrate that the X-ray light curves and measured neutron star surface temperatures are consistent with the expectations of the crust cooling paradigm. By using a thermal evolution code, we find that $\simeq$$1.2-3.2$ MeV~nucleon$^{-1}$ of shallow heat release describes the observational data well, depending on the assumed mass-accretion rate and temperature of the stellar core. We find no evidence for varying strengths of this shallow heating after different outbursts, but this could be due to limitations of the data. We argue that monitoring \source\ for up to $\simeq$1 year after future outbursts can be a powerful tool to break model degeneracies and solve open questions about the magnitude, depth and origin of shallow heating in neutron star crusts.
\end{abstract}

\begin{keywords}
accretion, accretion discs -- dense matter -- stars: neutron -- X-rays: binaries -- X-rays: individual (\source)
\end{keywords}


\section{Introduction}\label{sec:intro}

The study of transient accretion events in neutron star low-mass X-ray binaries (LMXBs) has proved to be a valuable exercise for gaining insight into the structure of neutron stars, and more generally the behaviour of matter at high densities \citep[e.g.,][for recent reviews]{wijnands2012,miller2013_NSreview,ozel2013_NSreview}. For example, the heating and subsequent cooling of the neutron star in response to an accretion outburst can be used to deduce valuable information about the thermal properties of its solid crust, as well as the superfluid properties of its liquid core \citep[e.g.,][]{rutledge2002,shternin07,brown08,page2013,turlione2013}.

During outburst, accretion of matter compresses the $\simeq$1-km thick neutron star crust. This compression induces electron captures in the outer crustal layers and pycnonuclear fusion reactions at several hundreds of meters depth, which together deposit a total energy of $\simeq$2 MeV per accreted nucleon in the crust \citep[e.g.,][]{haensel1990a,haensel2008,steiner2012}. As a result, the crust is heated out of thermal equilibrium with the core. This equilibrium is regained during the months--years after accretion stops, as the neutron star crust cools \citep[e.g.,][]{brown1998, rutledge2002, wijnands04_quasip, wijnands2012}. 

Crust cooling after an outburst has now been studied in detail for seven transient neutron star LMXBs: \ks\ \citep[e.g.,][]{wijnands2002,cackett2010}, \mxb\ \citep[e.g.,][]{wijnands2004,cackett2013_1659}, \xte\ \citep[e.g.,][]{fridriksson2011}, \exo\ \citep[e.g.,][]{diaztrigo2011,degenaar2014_exo3}, \igr\ \citep[e.g.,][]{degenaar2013_ter5}, \maxisource\ \citep[][]{homan2014}, and \terdrie\ \citep[][]{degenaar2015_ter5x3}. The observed cooling curves can be compared with thermal evolution calculations to gain detailed information on the neutron star such as the occurrence of ``chemical convection'' due to the phase separation between light and heavy nuclei in its outer layers \citep[e.g.,][]{degenaar2014_exo3,medin2014}, the presence of atomic nuclei with non-spherical shapes deep within the crust \citep[][]{horowitz2015}, and potentially even the surface gravity \citep[][]{deibel2015}. 

These studies have significantly improved our knowledge of neutron star crusts, but several open questions remain. One important issue is the apparent presence of an additional heat source in the upper region of the crust that cannot be explained by current nuclear heating models \citep[e.g.,][]{brown08,degenaar2011_terzan5_3}. So far, its origin remains unknown, and estimates for its magnitude vary from no shallow heating required in \xte\ and \terdrie\ \citep[][]{page2013,degenaar2015_ter5x3} up to $\simeq$10 MeV~nucleon$^{-1}$ in \maxisource\ \citep[][]{deibel2015}. 

Additional electron captures \citep[e.g.,][]{estrade2011} or nuclear fusion reactions \citep[e.g.,][]{horowitz2008} are likely not able to account for the apparent depth and magnitude of the additional heat \citep[][]{degenaar2013_xtej1709}. However, uncertainties on the nuclear symmetry energy could possibly allow the heat release from nuclear reactions to be up to $\simeq$2~MeV~nucleon$^{-1}$ higher than currently estimated \citep[][]{steiner2012}. This uncertainty could possibly lift the need for shallow heating for most crust-cooling sources studied so far, but the extreme case of \maxisource\ strongly suggests that additional heating mechanisms are operating, at least in some neutron stars \citep[][]{deibel2015}. Alternative explanations, not involving nuclear heating, include separation of light and heavy atomic nuclei that leads to a convective heat flux \citep[][]{horowitz2007,medin2011,degenaar2013_xtej1709}, release of orbital energy of the accreted material \citep[][]{deibel2015}, and excitation of gravitational modes in a differentially rotating layer on the neutron star \citep[][]{inogamov2010}. 

The need for extra heat extends to observations of thermonuclear X-ray bursts, which are bright flashes of X-ray emission resulting from unstable nuclear burning on the surface of neutron stars. For example, mHz quasi-periodic oscillations (QPOs) resulting from marginally-stable thermonuclear burning are observed at an inferred mass-accretion rate that is a factor $\simeq$10 lower than the theoretical predictions \citep[e.g.,][]{revnivtsev2001,altamirano2008,keek2009,linares2012_ter5_2}. This can be reconciled if there is an additional heat flux coming from the crust. Furthermore, the ignition of carbon that produces superbursts (very rare and energetic, hours-long X-ray bursts) can only be achieved if the crust temperature is considerably higher than accounted for by nuclear heating \citep[e.g.,][]{cumming06,keek2008_1608,altamirano2012}. Finally, the cessation of X-ray bursts as the mass-accretion rate increases seems to occur much more rapidly than can be accounted for by nuclear heating, and may also require additional energy release at shallow depth \citep[][]{zand2012}. Getting a better handle on this shallow heat release is thus an important step to improve our understanding of the observational properties of neutron stars, and the microphysics of their crust.

\subsection{\source}
\source\ is a transient LMXB that exhibits (normal) thermonuclear X-ray bursts and mHz QPOs \citep[e.g.,][]{koyama1981,revnivtsev2001,altamirano2008}. It is located at a distance of $D$$\simeq$5~kpc \citep[e.g.,][]{rutledge2001}, and spins at a frequency of $\nu_{\mathrm{s}}$$\simeq$550~Hz \citep[as inferred from coherent X-ray pulsations detected once during a $\simeq$150-s episode;][]{casella2008}. Of all neutron star transient LMXBs known, \source\ has one of the most active accretion histories, displaying outbursts of varying luminosity and duration roughly once every year \citep[e.g.,][]{kaluzienski1977,kitamoto1993,gungor2014_aqlx1}. Recently, \citet{campana2013} investigated the outburst properties of \source\ over a $\simeq$16-yr period (1996--2012), revealing 20 outbursts lasting from $t_{\mathrm{ob}}$$\simeq$1 to 26~weeks and with a luminosity ranging from $L_{\mathrm{X}}$$\simeq$$10^{35}$ to $10^{37}~\dist~\lum$. 

During quiescence, the X-ray luminosity of \source\ is $L_{\mathrm{X}}$$\simeq$$10^{33}$--$10^{34}~\dist~\lum$. Its quiescent spectrum consists of a soft, thermal component (dominating at energies $\lesssim$3~keV) and a harder component that can be described by a simple power law \citep[e.g.,][]{verbunt1994,rutledge2002_aqlX1,campana2003_aqlx1,cackett2011_aqlx1,campana2014}. The thermal emission component is thought to be coming from the neutron star surface, and is either due to it being heated by continued low-level accretion \citep[e.g.,][]{vanparadijs1987,zampieri1995,campana1997,deufel2001}, or due to radiation of heat deposited in its interior during accretion episodes \citep[e.g.,][]{brown1998,rutledge1999,campana2000_q}. 

The origin of the power-law spectral component is less well understood, but it may be related to the presence of a residual accretion flow and/or the magnetic field of the neutron star \citep[e.g.,][]{campana1998,rutledge2001,degenaar2012_amxp,chakrabarty2014_cenx4,wijnands2014}. Typically, this hard spectral component contributes $\simeq$50\% or less to the total unabsorbed 0.5--10 keV flux \citep[e.g.,][]{jonker2008,fridriksson2011,lowell2012,degenaar2013_ter5,degenaar2014_exo3,homan2014}. However, in exceptional cases it can account for all observed flux \citep[in \sax\ and \exoter; e.g., ][]{heinke2009,degenaar2012_1745}. Some neutron star LMXBs, including \source, display non-monotonic variability in quiescence, which seems to suggest that accretion may indeed persist down to very low levels \cite[e.g.,][]{campana1997,campana2004,rutledge2002,cackett2010_cenx4,cackett2013_cenx4,bernardini2013}. Often this variability is associated with the power-law spectral component, although in some cases the hard and soft emission components are varying in tandem \cite[e.g., in Cen X-4;][]{cackett2010_cenx4}. 

Two recent investigations have focussed on addressing the origin of the intensity variations observed from \source\ in quiescence. \citet{cackett2011_aqlx1} analysed 14 archival \chan\ and \xmm\ observations taken during the period 2000--2002 (two accretion outbursts were observed in this time frame). They found strong variations (up to a factor of 5) of the quiescent flux with time, although the data did not allow a conclusion on whether the variability was due to the thermal or power-law spectral component, or both. More recently, \citet{cotizelati2014} investigated the quiescent state of \source\ using frequent \swift/XRT monitoring over an 8-month interval in 2012. The source was found to be highly variable during this period, flaring up by a factor $>$10 on top of an overall decaying trend. Spectral analysis suggested that the observed variability could be accounted for by changes in the normalisation of the power-law component, although additional changes in the neutron star effective temperature could not be excluded.

In this work we investigate the hypothesis that thermal relaxation of the neutron star crust accounts for the long-term decay trend that is sometimes seen in the quiescent emission of \source. The source is bright enough to be detected in quiescence with the X-Ray Telescope \citep[XRT;][]{burrows05} onboard \swift\ \citep[][]{gehrels2004}, which has the flexibility to frequently monitor the source during its outburst decay and quiescent episodes. Within the crust cooling paradigm, a longer and brighter outburst should cause more intense heating of the crust and hence result in a higher quiescent temperature and longer cooling time scale than for a shorter and fainter outburst. To test if such trends are present, we investigated \swift/XRT data obtained during the decay and subsequent quiescence of three of the most recent outbursts of \source: 2011, 2013, and 2015.

\begin{figure}
 \begin{center}
\includegraphics[width=8.5cm]{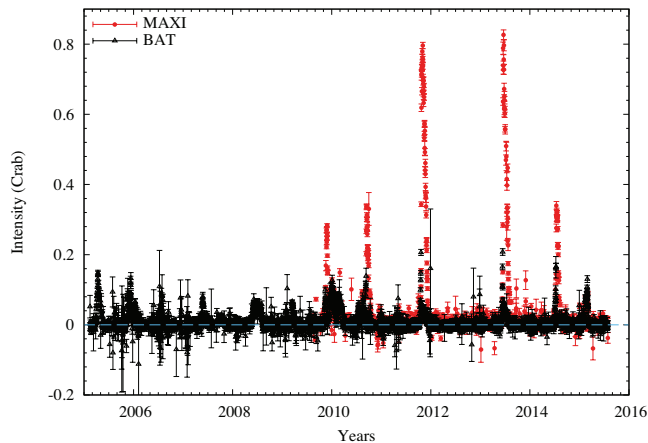}
    \end{center}
\caption[]{Long-term \maxi\ (2--20 keV; red) and \swift/BAT (15--50 keV; black) light curves of \source\ in Crab units (binned per day). Note that \maxi\ was launched in 2009. 
}
 \label{fig:lclong}
\end{figure}


\section{Observations and data reduction}\label{sec:obs}
Figure~\ref{fig:lclong} displays the outburst history of \source\ in the past 10 years, as registered by the transient monitoring programs of \swift/BAT \citep[15--50 keV;][]{krimm2013}\footnote{http://swift.gsfc.nasa.gov/results/transients/AqlX-1/} and \maxi\ \citep[2--20 keV;][]{maxi2009}.\footnote{http://maxi.riken.jp/top/index.php?cid=1\&jname=J1911+005} This illustrates the frequent activity displayed by this LMXB, and the differences in duration and (peak) intensity from outburst to outburst. The 2011 and 2013 outbursts were the brightest two detected from \source\ in the past decade.

With the aim of searching for crust cooling, we triggered a cycle 11 \swift\ proposal (PI: Degenaar) to monitor \source\ as it decayed to quiescence after its bright 2013 outburst. The program comprised a total of 25 observations of $\simeq$2~ks each, taken at a cadence of once every $\simeq$2--7 days, and spanning a time frame of 2013 July 24 to November 12 (obsID 32888001--25). At this time the source region entered the $\simeq$3-month long Sun constrained window, preventing further monitoring with \swift/XRT. The first seven observations (obsID 32888001--7) tracked the decay of the outburst, and were taken in windowed timing (WT) mode. In subsequent observations the count rate was sufficiently low to observe \source\ in photon counting (PC) mode.

\begin{table*}
\caption{Outburst properties from \swift/BAT and \maxi\ monitoring. \label{tab:ob}}
\begin{threeparttable}
\begin{tabular*}{0.96\textwidth}{@{\extracolsep{\fill}}lcccccccc}
\hline
Year & Start & End & $f_{\mathrm{BAT}}$ & $f_{\mathrm{MAXI}}$ & $F_{\mathrm{2-50}}$ & $L_{\mathrm{2-50}}$ & $\dot{M}$ & Reference for outburst start \\
& (MJD) & (MJD) & \multicolumn{2}{c}{($10^{-2}~\fluence$)} & ($\flux$) & ($\lum$) & ($\mdotg$) &  \\
\hline
2011 & 55845 & 55907 & $0.21$ & $6.0$ & $1.4\times10^{-8}$ & $4\times10^{37}$ & $4\times10^{17}$  &  \citet{yamaoka2011} \\
2013 & 56448 & 56513 & $0.20$ & $5.8$ & $1.3\times10^{-8}$ & $4\times10^{37}$ & $4\times10^{17}$ &  \citet{meshcheryakov2013_aqlx1}\\
2015 & 57069 & 57088 & $0.20$ & $0.28$ & $2.5\times10^{-9}$ & $7\times10^{36}$ & $8\times10^{16}$ &  \citet{ueno2015_aqlx1} \\
\hline
\end{tabular*}
\begin{tablenotes}
\item[]Note. -- The start time is based on the given references and the end time reflects the date when the source dropped below the sensitivity limit of \maxi. Count rates were converted into fluxes assuming a Crab-like spectrum to deduce the observed fluences for \swift/BAT (15--50 keV), $f_{\mathrm{BAT}}$, and \maxi\ (2--20 keV), $f_{\mathrm{MAXI}}$. $F_{\mathrm{2-50}}$ gives the estimated average flux in the 2--50 keV band from combining both instruments, and $L_{\mathrm{2-50}}$ is the corresponding average luminosity for $D=5$~kpc. $\dot{M}$ gives the estimated average mass-accretion rate during the outburst (see Section~\ref{subsec:ob}).
\end{tablenotes}
\end{threeparttable}
\end{table*}

Given the similarities of the 2013 accretion episode with the 2011 outburst (see Section~\ref{subsec:ob}), it is instructive to compare the 2013 decay and quiescence data with 41 XRT observations obtained in 2012, $\simeq$120--375~days after the 2011 outburst (obsID 31766040--81). These observations, all in PC mode, were obtained as part of a $\simeq$8-month long monitoring program where $\simeq$2--5~ks exposures were taken approximately every week between 2012 March 15 and November 9 \citep[][]{cotizelati2014}. There was also good \swift/XRT coverage of the decay and subsequent quiescence of the faint and short 2015 outburst of \source, which provides an interesting comparison with the 2011/2013 epochs. The 2015 data comprise of 37 observations of $\simeq$1~ks each (obsID 33665001--6 in WT mode, 33665007--38 in PC mode). The first 28 of these  were performed between 2015 March 8 and April 7, covering the outburst decay and first $\simeq$25~days of quiescence. After a $\simeq$3-month gap, a second set of 9 quiescent state observations was taken between 2015 June 29 and July 17. 

To extract data products we used the online XRT repository.\footnote{www.swift.ac.uk/user\_objects/} The main analysis steps are summarised below; details are described in \citet{evans2007} and \citet{evans2009}. Observations were typically obtained in WT mode when the XRT count rate was $\gtrsim$1$~\cnts$, and in PC mode when the source was fainter. Since the intensity of \source\ changes by orders of magnitude, a variable extraction size was applied to optimise the signal to noise ratio for all observations: When the source was bright we used a larger extraction region to maximise the number of counts measured, whereas at lower count rates the size was reduced to prevent that the collected counts were dominated by the background. For the PC data we used circular extraction regions with radii of $21''$--$71''$ to obtain source count rates and spectra, and a surrounding annulus with an inner/outer radius of of $142''$/$260''$ to extract background events. In WT mode, a box with a width of $71''$ was used for the source and background events were collected from the edges of the window, excluding $283''$ around the source. 

For observations with count rates of $>$$150~\cnts$ (WT) and $>$$0.6~\cnts$ (PC), pile-up corrections were applied by exercising as many inner pixels as needed for the count rate to drop below these limiting values \citep[][]{vaughan06,romano2006}. For creating count rate light curves, corrections for losses due to pile-up and dead zones on the CCD (hot pixels and bad columns) were applied by simulating the complete (i.e., unaffected) and partial (i.e., affected) PSFs for each interval, the ratio of which gives the required correction factor \citep[][]{evans2007}. Furthermore, the background count rates were scaled based on the ratio between the source and background extraction area.

To account for dead zones in the spectral analysis, an exposure map was created for each interval and used to produce an ancillary response file (arf) with \textsc{xrtmkarf}. The most up-to-date version of the response matrix file (rmf) was taken from the calibration data base (ver. 15). Differences in source and background extracting areas were accounted for by setting the \textsc{backscale} keyword in the spectral headers. During the outburst decay the count rate is high and varies rapidly, hence spectra extracted from single observations need to be analysed separately to study the flux and spectral shape evolution. However, in quiescence the count rate is much lower and evolves much slower (Section~\ref{subsec:decay}), so that multiple observations can and need to be combined to obtain sufficient counts for spectral analysis (Section~\ref{subsec:qspec}). In that case, weighted arfs and rmfs were created using \textsc{addarf} and \textsc{addrmf}. Prior to spectral fitting with \textsc{XSpec} \citep[ver. 12.8;][]{xspec}, spectra were grouped to contain a minimum number of 20 photons per bin. 

Throughout this work we assume a distance of $D$$=$5.0~kpc for \source\ \citep[see][for a discussion]{rutledge2001}. Furthermore, we adopt a neutron star mass of $M$$=$$1.6~\Msun$ and a radius of $R$$=$11~km for calculating mass-accretion rates (Section~\ref{subsec:ob}) and for our quiescent spectral analysis (Section~\ref{subsec:qspec}). This is to ensure self-consistency with our theoretical modelling (Section~\ref{subsec:coolmodel}). Errors presented in tables and plots reflect $1\sigma$ confidence intervals.

\begin{figure}
 \begin{center}
\includegraphics[width=8.5cm]{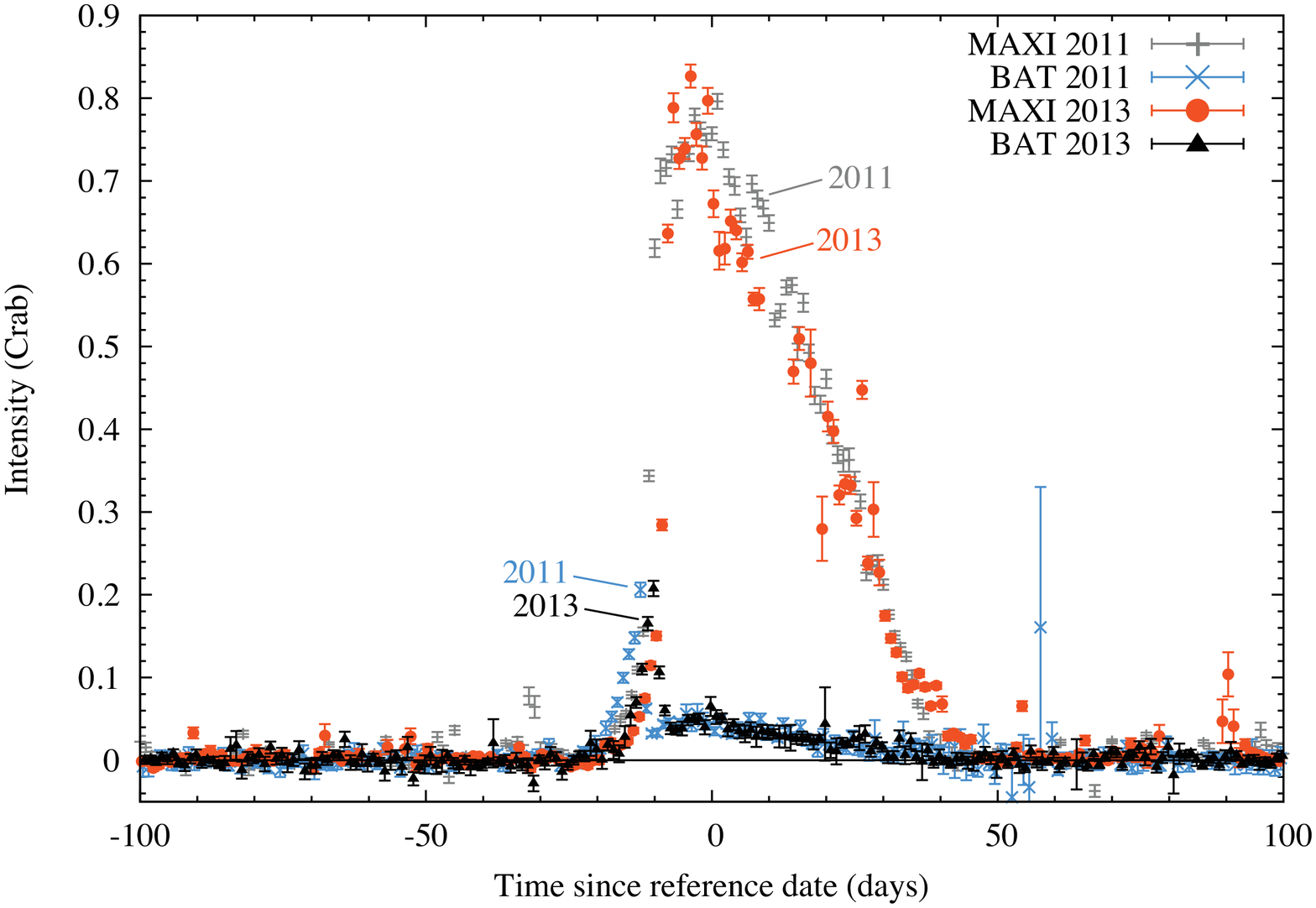}\vspace{+0.2cm}
\includegraphics[width=8.5cm]{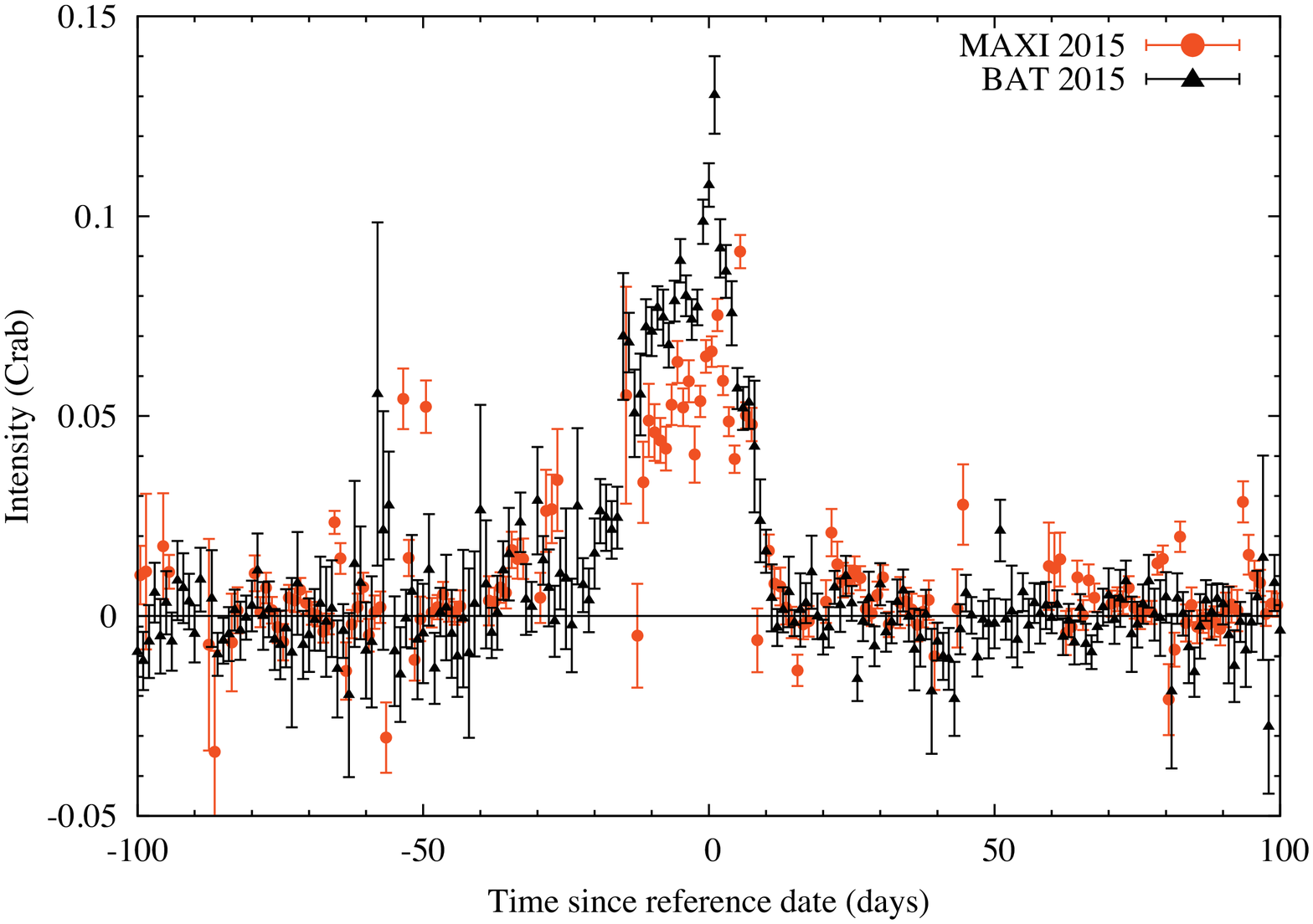}
    \end{center}
\caption[]{ \maxi\ (2--20 keV) and \swift/BAT (15--50 keV) light curves (binned per day), where different symbols/colours are used for different years/instruments. We note that the outburst profiles are not necessarily the same for the two instruments, because these track separate energy bands where different emission components dominate (e.g., the accretion disc is most prominent in the softer X-rays and the corona in the harder X-rays). {\bf Top}: The bright and long 2011 and 2013 outbursts overlaid, with the zero point on the time axis chosen near the peak of the respective outbursts. {\bf Bottom}: The short and faint 2015 outburst.
}
 \label{fig:maxibatlc201120132015}
\end{figure}


\section{Analysis and results}

\subsection{Outburst properties}\label{subsec:ob}
The thermal relaxation of a neutron star crust depends on the amount of heat that was generated during the preceding accretion episode. Therefore, to model crust cooling curves of \source, we need a handle on the duration and average luminosity of its various outbursts. For this purpose we examined the daily \maxi\ (2--20 keV) and \swift/BAT (15--50 keV) light curves of the 2011, 2013, and 2015 outbursts, which are shown in Figure~\ref{fig:maxibatlc201120132015}. We converted the \swift/BAT and \maxi\ count rates into fluxes by assuming a Crab-like spectrum, and then multiplied the average flux with the observed outburst duration to estimate the outburst fluence for each instrument. The Crab spectrum is on average an absorbed power-law with $\Gamma=2.1$ and $N_{\mathrm{H}} = 4 \times 10^{21}~\nh$ \citep[][]{kirsch2005}. If the photon index differs by $\simeq$1 (i.e., when the spectrum is harder or softer) or the absorption is a factor $\simeq$2 higher (see Section~\ref{subsec:qspec}), the 2--50~keV flux differs by $\simeq$25\%. Therefore this approach seems reasonable for \source. The main properties of the different outbursts are listed in Table~\ref{tab:ob}.

The 2011 and 2013 outbursts of \source\ both lasted for $\simeq$8 weeks and had very similar energetics (Table~\ref{tab:ob}). This is also illustrated by their very similar outburst profiles shown in Figure~\ref{fig:maxibatlc201120132015}, where the 2011 and 2013 light curves are overlaid. Comparing the fluences inferred from \swift/BAT and \maxi\ suggests that the energy release was concentrated toward softer X-ray fluxes. The outburst in 2015 was much shorter ($\simeq$3 weeks) and the average 2--50 keV flux was a factor $\simeq$5 lower than in 2011/2013. The fluence in the 15--50 keV band was not very different from that of the longer/brighter 2011 and 2013 outbursts, but the 2--20 keV fluence was a factor $\simeq$20 smaller in 2015 (Table~\ref{tab:ob}). This demonstrates the need for proper broad-band coverage to estimate outburst fluences.

Based on the energetics inferred from \swift/BAT and \maxi\ monitoring, we estimate the average mass-accretion rates during the various outbursts. When converting from the 2--10 keV band to bolometric flux, typically a correction factor of $\simeq$3 is assumed \citep[][]{zand07}. Since we have a broader bandpass of 2--50 keV, but with the flux concentrated towards lower energies (see above), we assumed $L_{\mathrm{bol}}$$\simeq$$2 \times L_{\mathrm{2-50}}$. We then compute $\dot{M}$$=$$RL_{\mathrm{bol}}/GM$, where $G$ is the gravitational constant and $R$$=$11~km and $M$$=$$1.6~\Msun$ are our assumed neutron star mass and radius (see Section~\ref{sec:obs}). Making such estimates is standard practice, but these provide only a rough approximation. Apart from assumptions about the outburst spectral shape (see above), there are considerable uncertainties in bolometric and anisotropy corrections when converting observed fluxes to bolometric luminosity, and also in the radiation efficiency when translating bolometric luminosity to mass accretion rate.

\begin{figure}
 \begin{center}
\includegraphics[width=8.5cm]{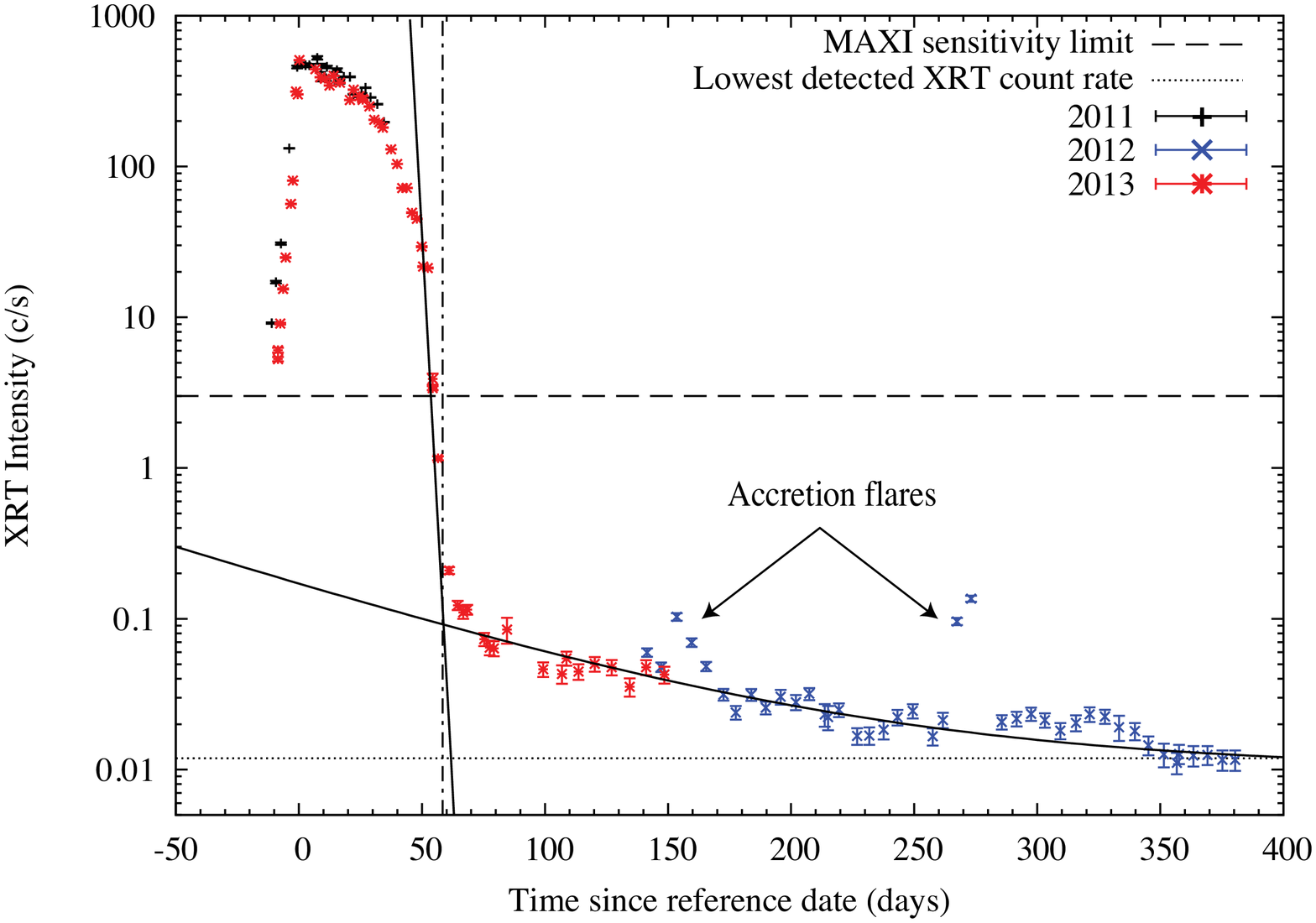}\vspace{0.3cm}
\includegraphics[width=8.3cm]{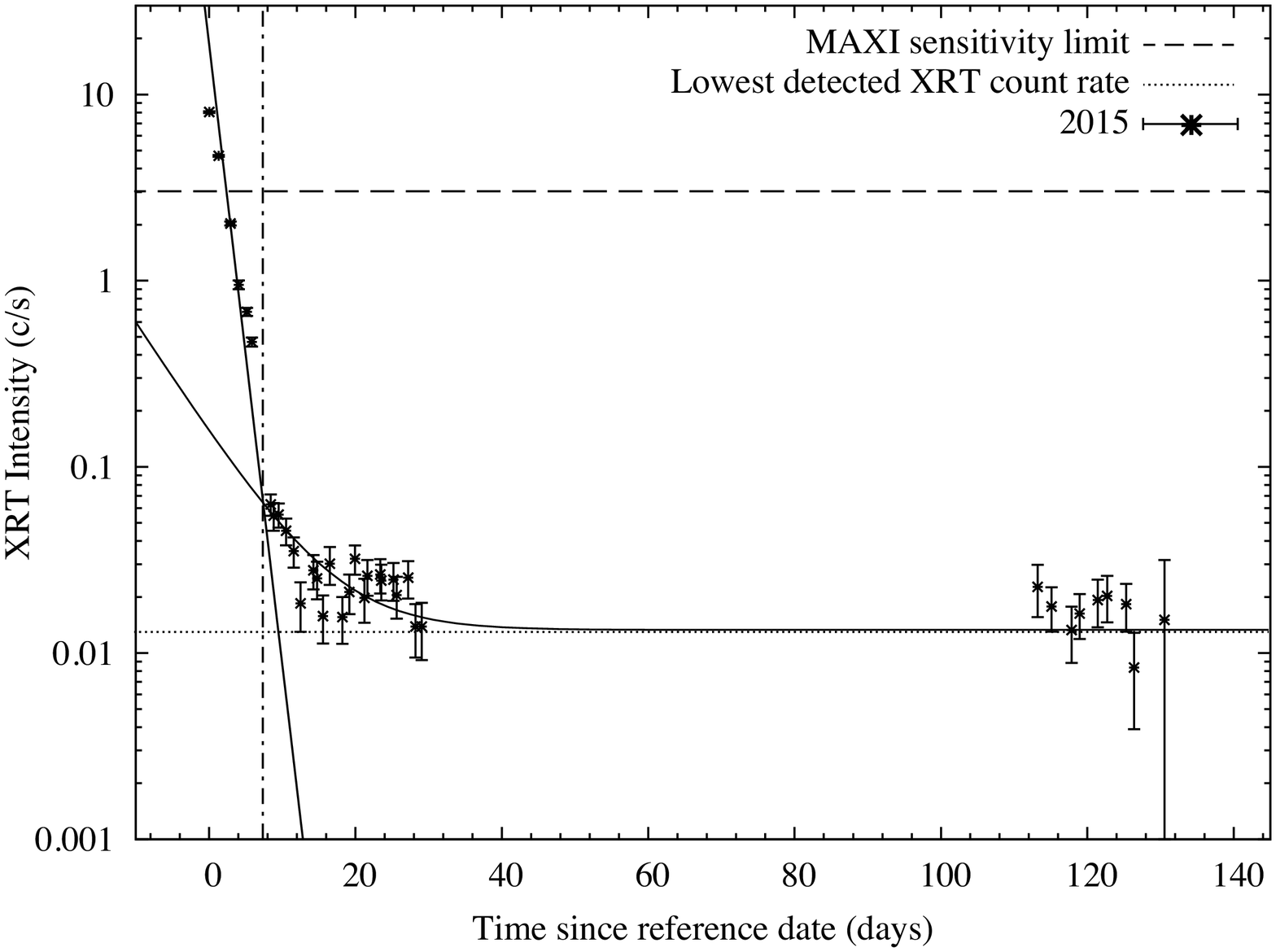}
    \end{center}
\caption[]{\swift/XRT count rate light curves of \source\ (0.5--10 keV, binned per observation). The horizontal lines indicate the (approximate) \maxi\ sensitivity limit (dashed) and the lowest count rate ever detected from \source\ with \swift/XRT (dotted). The latter was measured over 7 observations obtained during a $\simeq$30~day interval in 2012 ($>$1 yr after an outburst; Section~\ref{subsec:qspec}). Solid curves indicate exponential decay fits to different parts of the light curves, and the vertical dashed-dotted lines mark the transition from outburst decay to quiescence estimated from their interception. {\bf Top}: The 2011--2013 data overlaid, where the zero-point of the time axis was chosen near the peak of the respective outbursts: 2011 November 2 (MJD~55867), and 2013 June 17 (MJD~56460). {\bf Bottom}: The 2015 outburst decay and subsequent quiescent evolution in days since 2015 March 8 (MJD~57089).}
 \label{fig:xrtlc20112013}
\end{figure}

In 2012, there was no outburst detected from \source\ with \maxi\ and \swift/BAT (Figure~\ref{fig:lclong}). However, \swift/XRT monitoring did catch two brief, faint accretion flares (Figure~\ref{fig:xrtlc20112013} top). These were characterised by \citet{cotizelati2014} as having a duration of $\simeq$2--4 weeks, a peak luminosity of $L_{\mathrm{X}}$$\simeq$$10^{35}~\dist~\lum$, and an average mass-accretion rate of $\dot{M}$$\simeq$$3\times10^{14}~\mdotg$ (0.5--10 keV). Since the energy injected into the neutron star during such small accretion flares is much lower than that of the main outbursts, it does not lead to significant (i.e., detectable) heating of the crust \citep[e.g.,][]{turlione2013}.

\subsection{Decay properties}\label{subsec:decay}
From pointed \swift/XRT observations we can obtain valuable constraints on the outburst decay, beyond the detectability of all-sky monitors. As can be seen in Figure~\ref{fig:xrtlc20112013} (top), the entire 2013 outburst of \source\ was well-sampled by the XRT. This shows that after the source had dropped below the \maxi\ sensitivity threshold, the XRT count rate initially decayed rapidly by $\simeq$2 orders of magnitude in $\simeq$12~days, after which it decreased much more gradually over the next $\simeq$80~days until monitoring stopped due to Solar constraints. The count rate detected during the last few observations of 2013 was a factor $\simeq$4 higher than its lowest XRT intensity ever detected (in 2012 October--November; see Figure~\ref{fig:xrtlc20112013}).

Similar rapid declines followed by more gradual decays have been seen for several neutron star LMXBs when dense X-ray coverage was available during the final stage of an outburst \citep[e.g.,][]{fridriksson2010,campana2014,homan2014}. This is generally interpreted as the transition from outburst decay to cooling of the neutron star crust in quiescence. With this in mind we fitted the two decay parts in the \swift/XRT light curve of \source\ to simple exponential decay functions (solid lines in Figure~\ref{fig:xrtlc20112013}). For 2013, this yielded a characteristic decay time of $\tau_1$$=$$1.2\pm0.1$~day for the rapid decline, and $\tau_2$$=$$95.9 \pm 7.0$~day for the slower decay. From the intersection between these curves we estimate that quiescence started around 2013 August 14 (MJD~56518). This is $\simeq$58 days after the outburst peak measured by the XRT. 

In 2011, Solar constraints prevented XRT coverage of the full outburst and transition to quiescence (Figure~\ref{fig:xrtlc20112013}, top). However, the \swift/BAT and \maxi\ monitoring light curves of the 2011 and 2013 outbursts were strikingly similar (Figure~\ref{fig:maxibatlc201120132015} top). This is reinforced by XRT observations of the first part of the outbursts, which are overlaid in Figure~\ref{fig:xrtlc20112013} (top). We therefore estimate the time of the transition to quiescence by assuming similar outburst evolution and decay profiles. For the 2013 outburst, we estimated that \source\ hit quiescence $\simeq$58~days after the outburst peak. If we extrapolate this to the 2011 outburst, the peak of which was measured with the XRT on November 2 (MJD~55867), we estimate that the onset of quiescence occurred around 2011 December 30 (MJD~55925; Figure~\ref{fig:xrtlc20112013} top). It is of note that the quiescent data obtained after the two outbursts line up very smoothly, suggesting very similar quiescent behaviour as well (Figure~\ref{fig:xrtlc20112013} top). Combined with the similarities in outburst energetics, this implies that the different data sets can be combined to track the thermal evolution of the neutron star over $\simeq$300~days (Section~\ref{subsec:qspec} and~\ref{subsec:coolmodel}).

For the fainter, shorter 2015 outburst only the tail was observed with the XRT (Figure~\ref{fig:xrtlc20112013} bottom). Nevertheless the same general trend of a rapid decay followed by a slower decline is apparent: the count rate initially drops by $\simeq$2 orders of magnitude in $\simeq$8~days, but decreases much slower over the subsequent $\simeq$25~days. After a gap of three months, the source was detected at a similar count rate (with no strong variability among 9 observations performed over 20 days time), indicating no further decrease in intensity (Figure~\ref{fig:xrtlc20112013} bottom). For the 2015 outburst we find characteristic decay times of $\tau_1$$=$$1.6 \pm 0.1$~day and $\tau_2$$=$$7.1 \pm 1.2$~day for the rapid and slower decay, respectively. From the intersection, we determine that the transition to quiescence took place around 2015 March 18 (MJD$\simeq$57100; Figure~\ref{fig:xrtlc20112013} bottom). 

It is of note that despite the differences in outburst profiles and energetics (Section~\ref{subsec:ob}), the rapid declines are very similar for the 2013 and 2015 outbursts ($\tau$$\simeq$1--2~days). Decay times of $\tau$$\simeq$1.7--1.8~days were also reported for the 1997 and 2010 outbursts of \source\ \citep[][]{campana2014}. Such similarities may be expected if the rapid decay is caused by the onset of a propeller mechanism \citep[e.g.,][]{campana1998_aqlx1,campana2014,zhang1998,asai2013}. Alternatively, the rapid decay may correspond to draining of the accretion disc on a viscous time scale \citep[e.g.,][]{fridriksson2010,armas2013_2}. 

The decay time in quiescence is, however, much longer after the bright 2013 and 2011 outbursts than after the fainter 2015 one ($\simeq$100 vs $\simeq$7~days, respectively). The intensity also differs; after the 2015 outburst the count rate decayed to $\simeq$$0.02~\cnts$ in $\simeq$25~days, whereas 25~days after the 2013 outburst the count rate was $\simeq$$0.06~\cnts$, and in 2012 the count rate did not decay to $\simeq$$0.02~\cnts$ until $\simeq$200~days after entering quiescence (Figure~\ref{fig:xrtlc20112013}). Both the higher intensity and the slower decay seen after the 2011 and 2013 outbursts (compared to 2015) would be expected if cooling of the accretion-heated neutron star crust is observed. This is because long and bright outbursts such as seen in 2011 and 2013 should generate more heat and hence result in a longer cooling time scale than shorter and fainter ones such as in 2015 (see Section~\ref{subsec:coolmodel}).

\begin{figure}
 \begin{center}
\includegraphics[width=8.5cm]{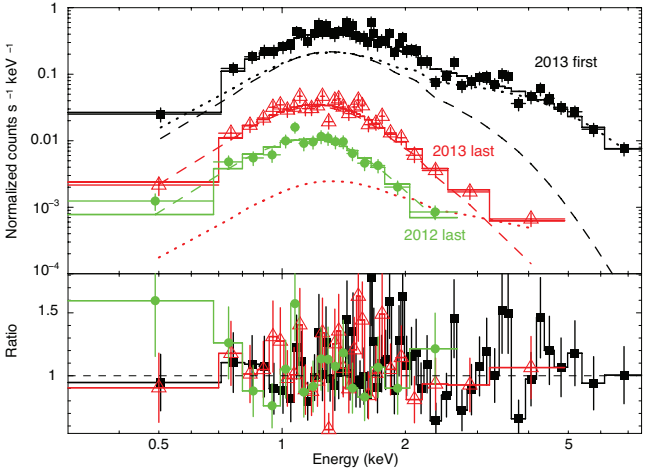}
    \end{center}
\caption[]{Illustrative \swift/XRT spectra together with fits to a combined neutron star atmosphere (dashed lines) and power law (dotted lines) model. The total model fit is indicated by the solid curves. The bottom panel shows the data to model ratios. The observations used in this plot are 32888008 (2013 decay), 32888017--25 (2013 last quiescence interval), and  31766074--81 (2012 last quiescence interval).}
 \label{fig:spec}
\end{figure}

\begin{table*}
\caption{Spectral analysis results for \swift/XRT observations of \source.\label{tab:spec}}
\begin{threeparttable}
\begin{tabular*}{1.01\textwidth}{@{\extracolsep{\fill}}lcccccccccc}
\hline
Year & MJD & ObsIDs & $kT^{\infty}$  & $F_{\mathrm{X}}$ & $F_{\mathrm{th,bol}}$ & $f_{\mathrm{th}}$ & $L_{\mathrm{X}}$ & $L_{\mathrm{th,bol}}$ & $\chi_{\nu}^2$ (dof) & $P_{\chi}$ \\
&  &  & (eV) & \multicolumn{2}{c}{($10^{-12}~\flux)$} & & \multicolumn{2}{c}{($10^{33}~\lum)$} & & (\%)\\
\hline
2012 (q) & 56004.3 & 31766040--41 & $ 123.2\pm4.3$ & $3.00\pm0.68$ &$2.34\pm0.52$ & $0.61\pm0.25$ & $8.97\pm2.03$ & $7.0\pm0.69$ & 1.10 (18) & 34 \\
2012 (f) & 56019.6 & 31766042--44 & $123.2\pm4.3 $ & $4.02\pm0.61$ & $2.34\pm0.50$ & $0.46\pm0.14$ & $12.0\pm0.2$ & $7.0\pm0.69$ & 1.15 (33) &  25 \\
2012 (q) & 56032.0 & 31766045--48 & $  110.0\pm3.0$ & $1.62\pm0.30$ &$1.41\pm0.14$ & $0.63\pm0.22$ & $4.85\pm0.88$ & $4.22\pm0.42$ & 0.84 (19) &  66  \\
2012 (q) & 56055.7 & 31766049--54 & $  112.6\pm2.7$ & $1.46\pm0.26$ & $1.55\pm0.14$ & $0.79\pm0.25$ & $4.37\pm0.78$ & $4.64\pm0.41$ & 1.09 (18) &   35 \\
2012 (q) & 56086.7 & 31766055--61 & $  104.7\pm2.3$ &  $1.09\pm0.18$ & $1.23\pm0.10$ & $0.80\pm0.24$ & $3.26\pm0.54$ & $3.68\pm0.29$ & 1.01 (20)  &  45 \\
2012 (f) & 56127.0 & 31766062--63 & $138.5\pm4.3$ &  $6.25\pm0.94$ & $3.63\pm0.40$ & $0.47\pm0.13$ & $18.7\pm0.3$ & $10.9\pm0.1$ & 1.02 (35) &  43  \\
2012 (q) & 56145.5  & 31766064--68  & $ 104.7\pm2.6$  & $1.20\pm0.23$ & $1.17\pm0.11$ & $0.71\pm0.25$ & $3.59\pm0.70$ & $3.50\pm0.34$ & 1.27 (17) &  20 \\
2012 (q) & 56175.6 & 31766069--73  & $ 107.3\pm3.0$  & $1.14\pm0.25$ & $1.26\pm0.13$ & $0.80\pm0.28$ & $3.41\pm0.75$ & $3.77\pm0.38$ & 1.12 (13) &  34 \\
2012 (q) & 56205.1 & 31766074--81  & $100.1\pm1.7 $  & $0.69\pm0.07$ & $1.00\pm0.31$ & 1 & $2.07\pm0.19$ & $2.99\pm0.20$ & 0.70 (15) & 79 \\
2013 (d) & 56516.5  & 32888008  & $209.3\pm6.0$  & $59.8\pm1.7$ & $27.0\pm1.9$ & $0.41\pm0.04$ & $179\pm5$ & $80.7\pm5.6$ & 1.01 (54)  & 45   \\
2013 (q) & 56521.0  & 32888009  & $182.2\pm3.7$ & $9.12\pm0.85$ & $10.2\pm1.1$ & 1 & $27.3\pm2.5$ & $30.5\pm2.5$ & 0.80 (15)  &  68 \\
2013 (q) & 56524.2 &  32888010--12 & $159.0\pm2.7$ & $5.13\pm0.38$ & $6.03\pm0.93$ & 1 & $15.3\pm1.2$ & $18.0\pm1.2$ & 1.44 (23) & 10  \\
2013 (q) & 56535.2 & 32888013--16  & $141.8\pm3.0$ & $3.09\pm0.34$ & $3.80\pm0.71$ & 1 & $9.24\pm1.0$ & $11.4\pm1.0$ & 1.43 (11) &  15 \\
2013 (q) & 56559.2  & 32888017--25 & $126.5\pm2.4$ & $2.38\pm0.37$ & $2.51\pm0.19$ & $0.84\pm0.23$ & $7.12\pm1.11$ & $7.51\pm1.10$ & 1.16 (27) &  26 \\
2015 (d) & 57093.6 & 32888004  & $186.8\pm39.1$  & $55.3\pm21.3$ & $13.2\pm7.3$ & $0.21\pm0.23$ & $165\pm61$ & $39.5\pm21.5$ & 0.75 (16) &  74 \\
2015 (d) & 57094.7 & 32888005  & $174.3\pm19.0$  & $28.4\pm9.1$ & $8.7\pm3.2$ & $0.27\pm0.22$ & $85.0\pm27.0$ & $26.1\pm1.0$ & 0.67 (19) &  85 \\
2015 (d) & 57095.3 & 32888006  & $155.0\pm16.6$  & $19.6\pm1.54$ & $6.5\pm2.3$ & $0.28\pm0.23$ & $58.6\pm18.1$ & $19.3\pm1.8$ & 1.54 (18) & 7  \\
2015 (q) & 57099.3 & 33665007--11 & $132.5\pm3.3$  & $2.40\pm0.30$ & $3.02\pm0.63$ & 1 & $7.18\pm0.90$ & $9.03\pm1.88$ & 1.42 (8) & 18  \\
2015 (q) & 57110.4  & 33665012--28 & $112.6\pm2.3$  & $1.15\pm0.15$ & $1.55\pm0.40$ & 1 & $3.44\pm0.34$ & $4.64\pm1.19$ & 1.30 (13) &  20 \\
2015 (q) & 57210.8 & 33665030--38  & $107.3\pm3.6$  & $1.00\pm0.19$ & $1.38\pm0.38$ & 1 & $2.99\pm0.56$ & $4.13\pm1.14$ & 0.21 (3) &  89 \\
\hline
\end{tabular*}
\begin{tablenotes}
\item[]Note. -- In parenthesis we indicate in which state \source\ was caught, where d=decay of the outburst, q=quiescence, and f=flare). The 2015 decay observations (ID 33665004--6) were obtained in WT mode, all other data in PC mode. The start of quiescence was estimated to be around MJD 55925 for 2012, MJD 56518 for 2013, and MJD 57100 for 2015. $F_{\mathrm{X}}$ is the total unabsorbed model flux in the 0.5--10 keV band and $F_{\mathrm{th,bol}}$ the flux in the thermal component extrapolated to 0.01--100 keV. The fractional contribution of the thermal component to the total unabsorbed 0.5--10 keV flux is given by $f_{\mathrm{th}}$ (if $f_{\mathrm{th}}=1$ the data were adequately fit by a thermal model without requiring the addition of a power-law spectral component). The total 0.5--10 keV and thermal 0.01--100 keV luminosities are given by $L_{\mathrm{X}}$ and $L_{\mathrm{th,bol}}$, respectively (assuming $D=5$~kpc). $P_{\chi}$ gives the $p$-value, which represents the probability that deviations between the model and the data are due to chance. The following parameters were fixed during the spectral fits: $N_{\mathrm{H}}$$=$$6.9 \times 10^{21}~\nh$, $M$$=$$1.6~\Msun$, $R$$=$$11$~km, $D$$=$$5$~kpc, $N_{\mathrm{nsatmos}}=1$, and $\Gamma$$=$$1.7$. Errors represent 1$\sigma$ confidence intervals.
\end{tablenotes}
\end{threeparttable}
\end{table*}

\subsection{Quiescent spectral analysis}\label{subsec:qspec}
In quiescence, the XRT count rate of \source\ is $\simeq$0.01--0.1$~\cnts$ (Figure~\ref{fig:xrtlc20112013}), implying that individual pointings  collected only a limited number of counts ($\lesssim$200 per observation). We therefore combined subsequent observations that had similar count rates. Aiming for spectra with $\simeq$500 counts in total, we ended up with a set of 7 quiescent spectra for the 2012 data, 4 for 2013, and 3 for 2015. 

As reported by \citet{cotizelati2014}, the 2012 data included two $\simeq$2--4 weeks episodes during which the luminosity rose by a factor $>$10 (Figure~\ref{fig:xrtlc20112013} top), and the fractional contribution of the power-law spectral component increased. These flares were very likely caused by a brief spurt of accretion. Figure~\ref{fig:xrtlc20112013} (top) suggests that the first two observations of 2012 lie very close to the overall decay trend, but that the next three observations (obsID 31766042--44) lie above it. We therefore assumed that these three observations were taken during an accretion flare and extracted an average spectrum. Similarly, we extracted an average spectrum for two observations (obsID 31766062--63) that made up a second flare. 

There are a few observations with count rates of $\simeq$$1~\cnts$ during the decay of the 2013 and 2015 outbursts that occur very close in time to our estimated transition to quiescence, and can be described with a similar spectral model as for the quiescent state (see below). It is very likely that accretion is ongoing during those observations, but if the neutron star crust is very hot it is possible that cooling of the crust, rather than the surface temperature decreasing with falling mass-accretion rate, dominates the temperature evolution. To probe if this can indeed be the case, we include these decay points in our analysis (Table~\ref{tab:spec}).

We use \textsc{nsatmos} \citep[][]{heinke2006} to model the thermal emission from the neutron star surface, and \textsc{pegpwrlw} to describe any possible non-thermal emission tail. We include the effects of interstellar extinction by using the \textsc{tbabs} model with the \textsc{vern} cross-sections and \textsc{wilm} abundances \citep[][]{verner1996,wilms2000}. We assumed the same hydrogen column density for all data sets, since prior studies of \source\ in quiescence showed that this parameter is not changing between different epochs \citep[e.g.,][]{cackett2011_aqlx1,cotizelati2014,campana2014}. From a combined fit to all our \swift/XRT spectra we obtained $N_{\mathrm{H}}$$=$$(6.9\pm 0.2)\times10^{21}~\nh$.

For the \textsc{pegpwrlw} model component we set the energy boundaries to 0.5 and 10 keV, so that its normalisation represents the unabsorbed flux in this band. For the \textsc{nsatmos} component we fix $M$$=$$1.6~\Msun$ and $R$$=$11~km (see Section~\ref{sec:obs}). Furthermore, we fix the normalisation of this model to unity, which implies that we assume that the entire neutron star surface is emitting at all times. \citet{cotizelati2014} analysed the 2012 XRT quiescent observations and showed that the power-law normalisation had to be varying, whereas variations in the neutron star temperature were statistically not required but could not be excluded either. Since in the present work we aim to test the hypothesis that crust cooling can be observed from \source, we assume that both the \textsc{pegpwrlw} normalisation and the \textsc{nsatmos} temperature are variable. 

The data quality does not allow us to fit the power-law index for all individual observations. It is not known whether the power-law slope should change between outburst decay and quiescent states in neutron star LMXBs \citep[e.g.,][]{wijnands2014}. Simultaneously fitting our entire sample of spectra with the power-law index tied, yields a good fit with $\Gamma$$=$$1.7\pm0.1$ ($\chi_{\nu}^2=1.03$ for 460 dof). Allowing $\Gamma$ to be different for the decay and quiescent observations yields different values for the two states ($\Gamma$$=$$2.1\pm0.1$ and $\Gamma$$=$$0.8\pm0.5$, respectively). We suspect that these differences are due to the limited data quality. For instance, during the decay observations the power-law may be fitting part of the thermal emission \citep[see e.g.,][]{armas2011}, and in quiescence only few photons are detected at energies $>$3~keV to constrain the power-law index (see Figure~\ref{fig:spec}). Indeed,  allowing this parameter to vary does not provide a significant statistical improvement over fixing it for all spectra ($\chi_{\nu}^2=1.02$ for 428 dof, with an $F$-test probability of 0.19). A recent \chan\ and \xmm\ study of the decay of the 2010 outburst and the first $\simeq$30~days of quiescence also showed that the data was consistent with a single value over $\simeq$2 orders of magnitude in luminosity \cite[$\Gamma$$=$1.7;][]{campana2014}. In the present study we therefore assume that $\Gamma$$=$1.7 at all times.

We then fitted all individual spectra with $N_{\mathrm{H}}$, $M$, $R$, $D$, $\Gamma$ fixed at the above mentioned values, rendering only the neutron star temperature and power-law normalisation as free fit parameters. Once a fit converged, we used the \textsc{cflux} convolution model to determine the total unabsorbed 0.5--10 keV flux, the thermal flux in this band, and the 0.01--100 keV flux of the thermal component. We found that 7 of the spectra could be adequately described by a thermal emission model alone ($p$-probability of $P_{\chi}$$>$5\%), and did not require the addition of a hard emission tail (we obtained $F$-test values of $\gtrsim$$1\times10^{-3}$, indicating a large probability that the improvement of adding the power-law component was due to chance). The fit statistics of the remaining spectra improved significantly by the addition of the power-law component. The results of our spectral analysis are summarised in Table~\ref{tab:spec}. In Figure~\ref{fig:spec} we show example spectra to illustrate the data quality and the observed differences. 

\source\ is detected over a 0.5--10 keV luminosity range of $L_{X}$$\simeq$$(0.2-3)\times 10^{34}~\dist~\lum$ in the quiescent state observations. The average luminosity inferred for the two 2012 accretion flares is $L_{X}$$\simeq$$(1-2) \times 10^{34}~\dist~\lum$, whereas during the decay of the 2013 and 2015 outbursts we measure $L_{X}$$\simeq$$(6-18) \times 10^{34}~\dist~\lum$. The quiescent state observations are dominated by the thermal emission component, which contributes a fraction $f_{\mathrm{th}}\simeq$0.6--1 to the total unabsorbed flux in the 0.5--10 keV band. During the 2012 flares, and the 2013 and 2015 decay observations, this fraction is systematically lower, $f_{\mathrm{th}}\simeq$0.1--0.5 (Table~\ref{tab:spec}). The power-law spectral component is thus more prominent when accretion occurs.

We infer neutron star temperatures of $kT^{\infty}$$\simeq$$100-182$~eV in quiescence, $kT^{\infty}$$\simeq$$123-139$~eV for the accretion flares, and $kT^{\infty}$$\simeq$$155-209$~eV for the outburst decay observations. We observe an overall trend of the temperature decreasing with time after an outburst (Table~\ref{tab:spec}), as would be expected if the neutron star crust was heated during outburst and cooling in quiescence. The lowest temperature measured in our data set is $kT^{\infty}$$\simeq$$100$~eV and this was during the last set of 7 observations of 2012, which were performed over a $\simeq$30~day interval covering $\simeq$285--315 days after the 2011 outburst (Figure~\ref{fig:xrtlc20112013}). There are no lower temperatures reported in the literature; \citet{cackett2011_aqlx1} found a range of $kT^{\infty}$$\simeq$$106-142$~eV over many different quiescent epochs, and \citet{campana2014} reported $kT^{\infty}$$\simeq$$115-130$~eV after the 2010 outburst ($kT^{\infty}$$\simeq$$150-195$~eV during the decay).\footnote{Some caution is needed in comparing temperatures among different studies, since other values for $M$ and/or $R$ cause small systematic shifts in inferred neutron star temperatures; e.g., we obtain $kT^{\infty}$$\simeq$$104-219$~eV for our data set when using $M=1.4~\Msun$ and $R=10$~km.}   

To probe the effect of our choice of power-law index, we repeated our analysis for $\Gamma=1.0$ and~$2.5$ \citep[covering the range of values reported in the literature;][]{cackett2011,cotizelati2014}. For the observations that contained a significant power-law spectral component, this yielded neutron star temperatures shifted by a few eV compared to our results listed in Table~\ref{tab:spec}. However, the measurements were consistent within the 1$\sigma$ errors for all values of $\Gamma$ used. Moreover, an overall decaying trend is seen in the neutron star temperature as time progresses after an outburst, regardless of the choice of power-law index. Our general conclusions therefore do not rely on the choice of $\Gamma$.

\subsection{Crust cooling simulations}\label{subsec:coolmodel}

\begin{table*}
\caption{Model parameters for thermal evolution calculations of \source. \label{tab:model}}
\begin{threeparttable}
\begin{tabular*}{0.99\textwidth}{@{\extracolsep{\fill}}lccccccccc}
\hline
Model & Data set & Description & $\dot{M}$ & $t_{\mathrm{ob}}$ & $T_{\mathrm{core}}$ & $Q_{\mathrm{shallow}}$ & $d_{\mathrm{shallow}}$ & $\chi^2_{\nu}$ (dof) & $P_{\chi}$ \\
 &  & & (g~s$^{-1}$) & (yr) & (K) & (MeV~nucleon$^{-1}$) & & & (\%) \\ 
\hline
1 & 2013 & Basic & $4\times10^{17}$  & $0.15$  & $1.75\times10^8$ fix & $2.7\pm0.2$ & $27.7\pm0.3$ & 2.6 (2) & 7 \\
2 & 2013 & Including outburst decay & $4\times10^{17}$  & $0.15$  & $1.75\times10^8$ fix & $2.6\pm0.2$ & $27.2\pm0.2$ & 2.8 (3) & 4  \\
3 & 2012 & Basic  & $4\times10^{17}$  & $0.15$  & $1.75\times10^8$ fix  & $1.6\pm0.6$ & $27.5$ fix & 1.9 (5) & 9   \\
4 & 2012 & Free $T_{\mathrm{core}}$ & $4\times10^{17}$  & $0.15$  & $(1.6\pm0.1)\times10^8$ & $2.4\pm0.5$ & $27.5$ fix & 1.6 (4) & 17 \\
5 & 2012+2013 & Basic & $4\times10^{17}$  & $0.15$  & $1.75\times10^8$ fix  & $2.5\pm0.2$ & $27.5\pm0.3$ & 2.7 (9) & 0.3   \\
6 & 2012+2013 & Free $T_{\mathrm{core}}$ & $4\times10^{17}$  &  $0.15$  & $(1.5\pm0.1)\times10^8$ &  $3.2\pm0.2$ &  $27.9\pm0.2$ & 1.9 (8) & 6  \\ 
7 & 2012+2013 & Higher $\dot{M}$  & $8\times10^{17}$ & $0.15$  & $1.75\times10^8$ fix  & $1.2\pm0.1$ & $27.5\pm0.3$ & 3.0 (9) & 0.1  \\
8 & 2015 & Heating as per model 5 & $8\times10^{16}$  & $0.06$  & $1.75\times10^8$ fix  & $2.5$ fix & $27.5$ fix & 1.4 (3) & 24 \\
9 & 2015 & Heating as per model 2 & $8\times10^{16}$  & $0.06$  & $1.75\times10^8$ fix  & $2.6$ fix & $27.2$ fix & 2.0 (3) & 12 \\
\hline
\end{tabular*}
\begin{tablenotes}
\item[]Note. -- The following parameters were fixed at the quoted values during the model calculations: the average mass accretion rate during outburst, $\dot{M}$, and the outburst duration, $t_{\mathrm{ob}}$. All calculations assumed an impurity parameter of $Q_{\mathrm{imp}}=1$. For the 2013 data we first calculated a model without the decay data point (model 1) and then one with this point included (model 2). The magnitude of extra shallow heating, $Q_{\mathrm{shallow}}$, and the maximum depth of this shallow heat source, $d_{\mathrm{shallow}}$ (expressed as the log of the pressure $P$), were both free to vary except for the 2012 data, where lack of early cooling points did not allow to constrain $d_{\mathrm{shallow}}$ and therefore this parameter was fixed in models 3 and 4. The core temperature, $T_{\mathrm{core}}$, was fixed to the lowest temperature ever reported for \source, except for models 4 (2012 data) and 6 (2012+2013 data) in which it was free to vary. For the combined 2012+2013 data set we also calculated a model curve with a higher mass-accretion rate (models 7). The 2015 data could not be fit with the cooling model due to the limited number of data points. Therefore we calculated two ``predicted'' cooling curves (models 8 and 9) based on the estimated duration and mass-accretion rate for 2015, but using the microphysics from models 5 and 2, respectively (see Section~\ref{subsec:coolmodel} for details). The last two columns provide a quality measure of the model calculations. Quoted errors refer to 2$\sigma$ confidence intervals.
\end{tablenotes}
\end{threeparttable}
\end{table*}

\subsubsection{Numerical code and model fitting}
To investigate whether the decaying trend seen in both the XRT count rate light curves (Figure~\ref{fig:xrtlc20112013}) and the inferred neutron star temperatures (Table~\ref{tab:spec}) can plausibly be due to cooling of the accretion-heated crust, we confronted the observations with a thermal evolution code \textsc{dStar}.\footnote{The code is available at https://github.cam/nworbdc/dStar.} It solves the time-dependent equations for the evolution of temperature and luminosity via method-of-lines: it finite-differences the spatial derivatives, which yields a set of coupled ordinary differential equations in time for each spatial grid point \citep[see][]{brown08}. The finite-difference scheme explicitly conserves flux. Details of the full microphysics treatment can be found in \citet{brown08}; here we describe only the aspects most relevant to the current study. 

The neutron star ocean temperature is mapped to the photosphere temperature using separately computed models of the envelope. We used an envelope composed of $^4$He and $^{56}$Fe, following \citet{brown08}. Simulations were performed with the atmosphere temperature free to evolve as accretion proceeded. The neutron star has a set mass and radius, for which we chose $M$$=$$1.6~\Msun$ and $R$$=$11~km, similar to that assumed in several other theoretical crust cooling studies \citep[e.g.,][]{brown08,degenaar2014_exo3,medin2014}. The motivation is that neutron stars in LMXBs have likely accreted significant mass over their lifetime and may thus be more massive than the ``canonical'' $M$$=$$1.4~\Msun$. The often used Akmal-Pandharipande-Ravenhall (APR) equation of state corresponds to $R$$=$11~km for $M$$=$$1.6~\Msun$ \citep[][]{akmal1998}. 

The main input parameters for the model calculations are the outburst duration $t_{\mathrm{ob}}$ and mass-accretion rate $\dot{M}$, the level of impurities in the ion lattice which sets the thermal conductivity of the inner crust and is parametrized as $Q_{\mathrm{imp}}$, the temperature of the core $T_{\mathrm{core}}$, and finally the magnitude $Q_{\mathrm{shallow}}$ and maximum depth $d_{\mathrm{shallow}}$ of any shallow heat source. 

We fixed the outburst parameters to the values estimated from daily \maxi\ and \swift/BAT monitoring (Section~\ref{subsec:ob}). We found that the cooling curves calculated for \source\ are not sensitive to $Q_{\mathrm{imp}}$. This is because for outbursts with $t_{\mathrm{ob}}$$\lesssim$1~year, only the outer crust is heated \citep[e.g.,][]{page2013}. The thermal conductivity in the outer crust is set by electron-ion scatterings rather than electron-impurity scatterings, and is thus insensitive to $Q_{\mathrm{imp}}$ \citep[][]{brown08}. We therefore fixed $Q_{\mathrm{imp}}$$=$1 for all simulations, consistent with results found for other neutron stars \citep[e.g.,][]{brown08}, and predictions from molecular dynamics simulations \citep[][]{horowitz2007}. 

Since the thermal state of the core is not expected to change over a small number of accretion outbursts \citep[e.g.,][]{brown1998,rutledge2002}, we treat the core as a boundary condition with a fixed temperature. To determine the core temperature that matches the observational data for our chosen atmosphere composition, we ran a model with the mass-accretion rate set to zero. This way we found that a core temperature of $T_{\mathrm{core}}$$=$$1.75\times10^8$~K matches the lowest observed temperature ever reported for \source\ ($kT_{\mathrm{obs}}^{\infty}$$\simeq$100~eV; Section~\ref{subsec:qspec}). We initially fixed the core temperature at this value, but later we also explored model runs with the core temperature free to fit the data.

\begin{figure*}
 \begin{center}
\includegraphics[width=8.5cm]{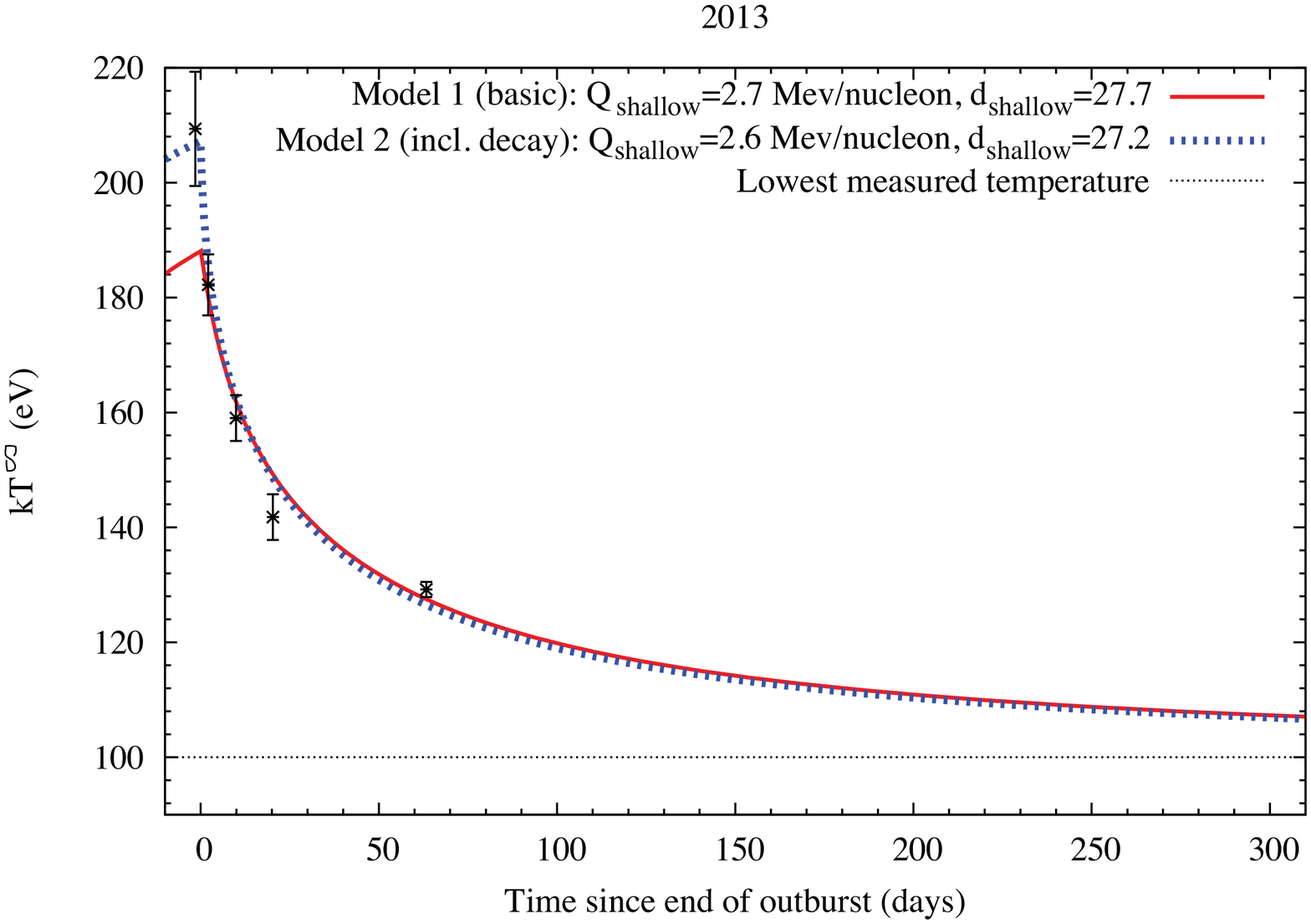}\hspace{0.5cm}
\includegraphics[width=8.5cm]{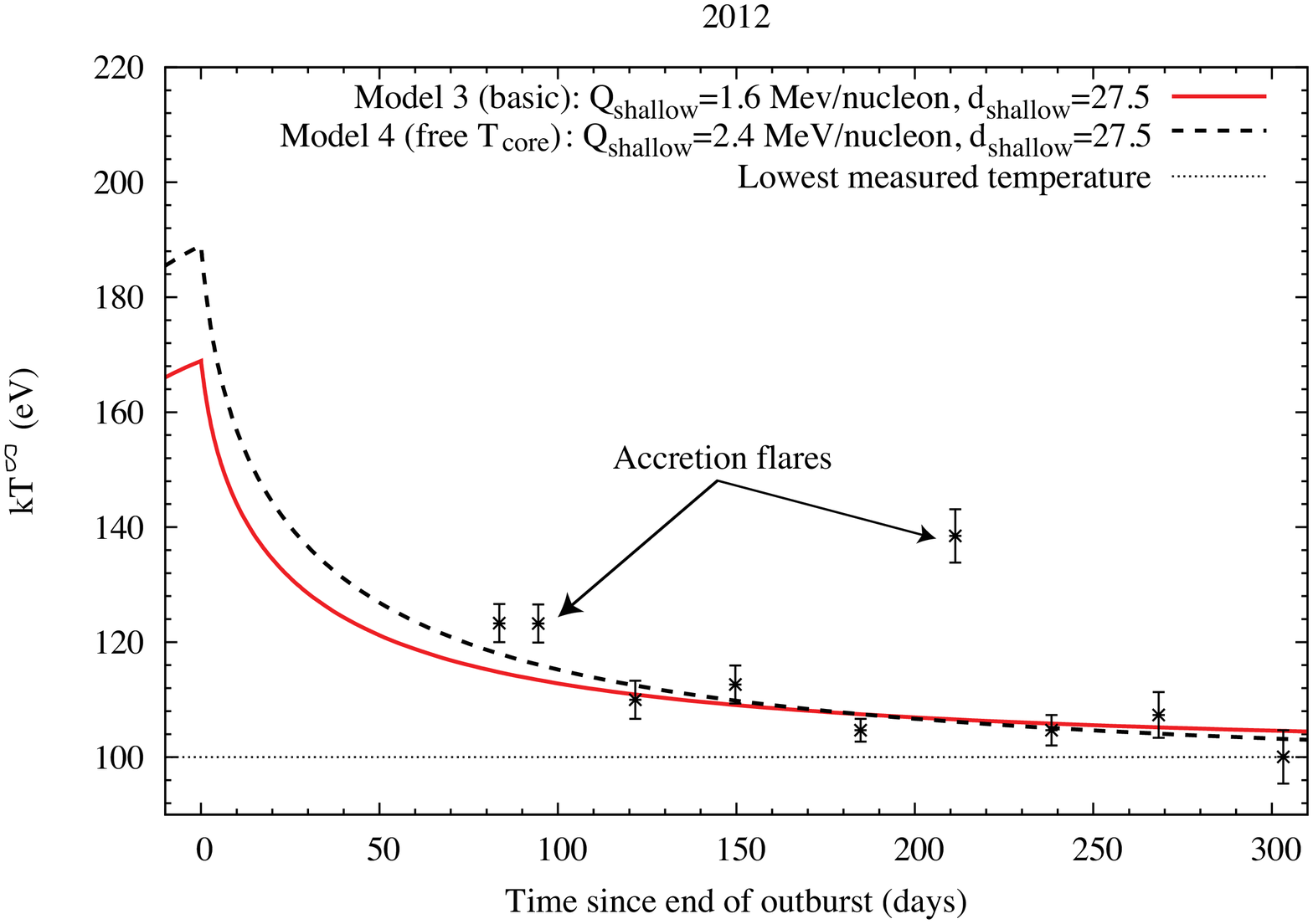}\vspace{0.5cm}
\includegraphics[width=8.5cm]{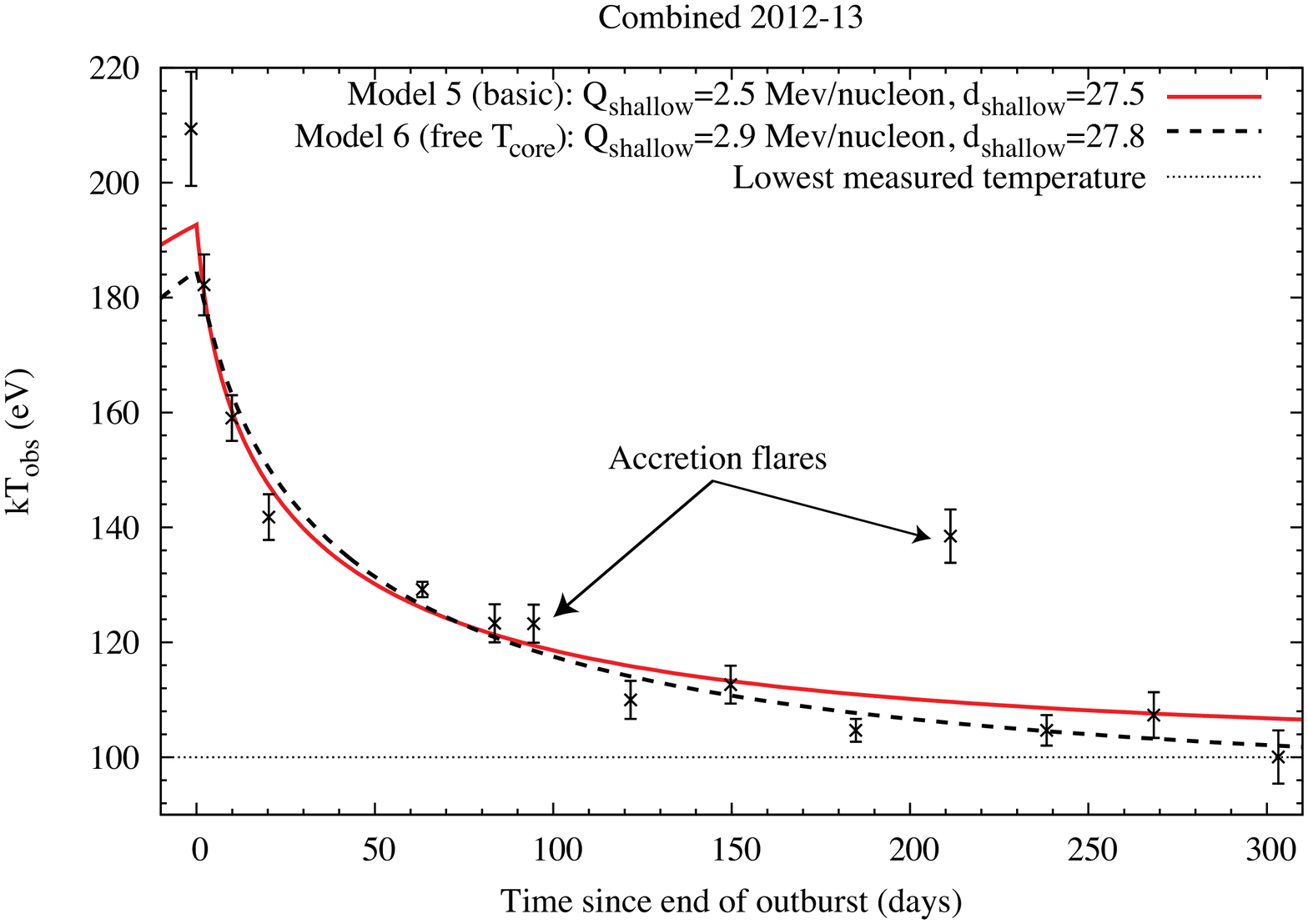}\hspace{0.5cm}
\includegraphics[width=8.5cm]{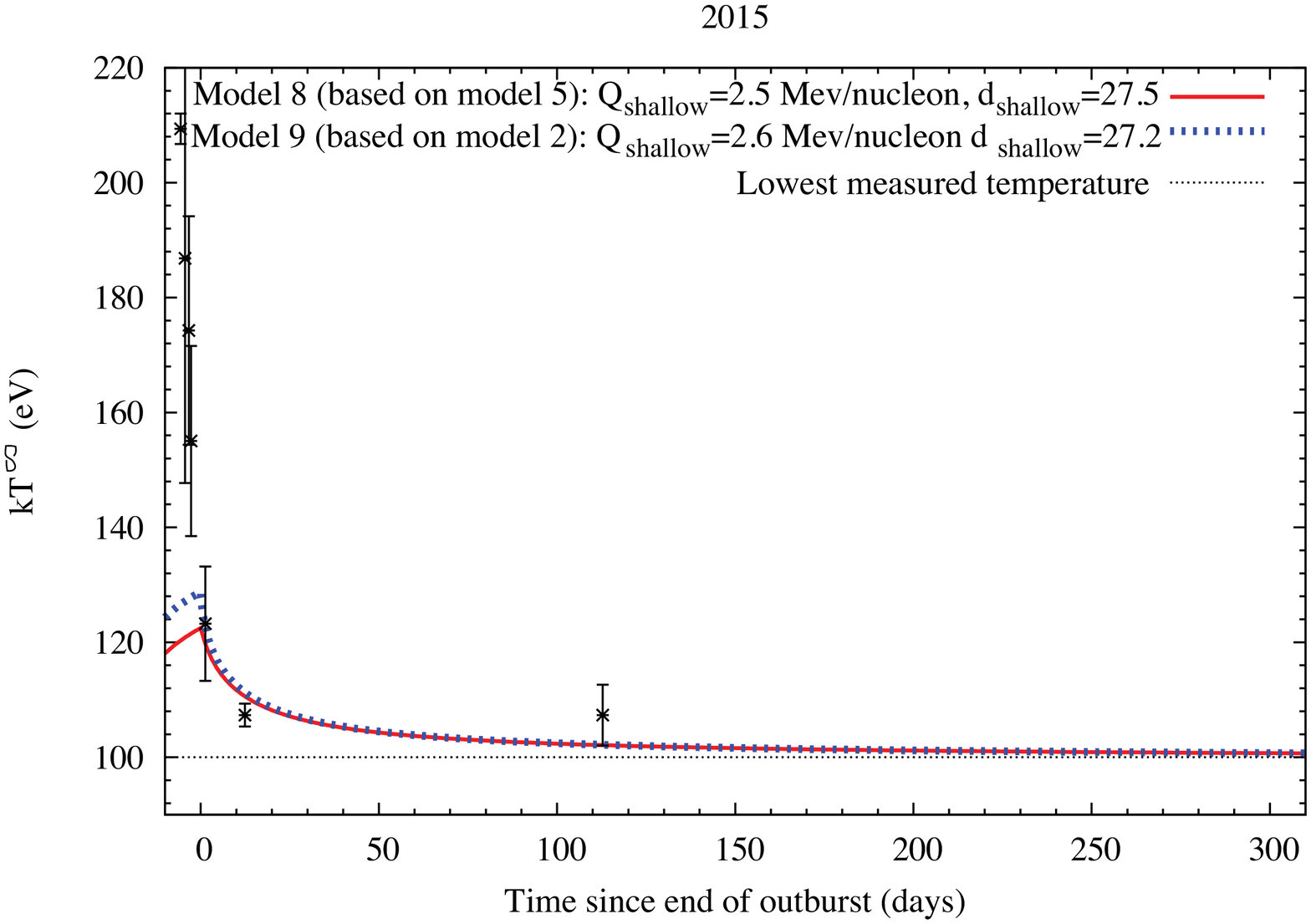}
	\end{center}
\caption[]{{Neutron star effective temperatures from analysis of \swift/XRT data, and model cooling curves (labelled according to Table~\ref{tab:model}). The shallow heating parameters corresponding to the model curves are printed in the plots. The two flares occurring in 2012 are shown for reference, but these were not included in the model fits. The end of the outburst was assumed to be MJD 55925 for the 2012 data, MJD 56518 for 2013, and MJD 57100 for 2015. {\bf Top left:} 2013 data and model fit with (blue dotted curve) and without (red) inclusion of the decay data point. {\bf Top right:} 2012 data and model fits for a fixed core temperature (red curve) and the core temperature free (black dashed). These data did not constrain the depth of the shallow heat, so it was fixed to $d_{\mathrm{shallow}} = 27.5$. {\bf Bottom left:} Combined 2013 and 2012 dataset with our basic model fit (red curve) and one with the core temperature free (black dashed). {\bf Bottom right:} There were not enough points to fit the 2015 data, so instead we calculated predicted cooling curves based on the 2015 outburst energetics and the shallow heating inferred for the 2012--2013 data. In red a calculation based on the 2012--2013 basic model, and in blue one for a lower depth of the shallow heat source as inferred for the 2013 data with the decay point included. The first three data points, inferred from the decay observations, are too high to be reproduced by our thermal evolution model. This likely indicates that these were dominated by accretion onto the stellar surface rather than crust cooling.
}}
 \label{fig:cool}
\end{figure*}

Our main interest is to constrain any possible shallow heating in \source, since this is currently one of the main open questions in this research area. In the model the depth in the crust is parametrized as $\log{P}$, where $P$ is the pressure [$\mathrm{erg~cm}^{-3}$]. The corresponding column depth is $y=P/g$ [$\mathrm{g~cm}^{-2}$], where $g$ is the gravitational acceleration. The boundaries of the crust are defined as $\log{P}$$=$27--32.5. The shallow heat source is implemented such that the heating rate per unit mass is uniform between the top of the crust ($\log{P}$$=$27), and the maximum depth of shallow heating ($\log{P}$$=$$d_{\mathrm{shallow}}$). The heating was assumed to be uniform because this is easiest to implement, and at present there is no information indicating that the heating should be non-uniform.

We thus calculated crust cooling models for \source\ using its estimated outburst duration and mass-accretion rate, $Q_{\mathrm{imp}}$$=$1 fixed, and the core temperature set to $T_{\mathrm{core}}$$=$$1.75\times10^8$~K. We then manually varied $Q_{\mathrm{shallow}}$ and $d_{\mathrm{shallow}}$, and evaluated the quality of a model calculation using the $\chi^2$ method. This was calculated by the standard method: For each temperature measurement the squared differences between the observed value and the model value were summed, where the differences were weighted by the inverse of the error on the corresponding data point. 

The parameter space was first surveyed with a low level of precision to obtain the approximate location of the global minimum in $\chi^2$. The surrounding parameter space was then explored with greater precision, taking steps of $\pm0.05$~MeV~nucleon$^{-1}$ in $Q_{\mathrm{shallow}}$, and $\pm0.05$ in $d_{\mathrm{shallow}}$. Once we obtained a good fit to the data, we probed the allowable range on the parameter values by changing $Q_{\mathrm{shallow}}$ or $d_{\mathrm{shallow}}$ until reaching a fit statistic of $\Delta \chi^2=4$ compared to the original fit. This way we set $\simeq$2$\sigma$ limits on the fit parameters. In a later stage we also ran model calculations with the core temperature free to vary. For those calculations we performed a similar grid search to determine a $\simeq$2$\sigma$ confidence interval for the value of the core temperature.

The input used for our different model calculations is listed in Table~\ref{tab:model}, and the results are shown in Figure~\ref{fig:cool}. For reference we show the temperatures measured during two accretion flares in Figure~\ref{fig:cool}, but these two points were not included in our model fits, because accretion onto the neutron star surface likely dominates the temperature evolution during these flares (Section~\ref{subsec:qspec}).

\subsubsection{Numerical models for the 2011--2013 data}
For 2013, we find that a shallow heat source with a magnitude of $Q_{\mathrm{shallow}}$$=$2.7~MeV~nucleon$^{-1}$ produces a good match to the  data (model 1 in Table~\ref{tab:model}; Figure~\ref{fig:cool} top left). For 2012, we obtain a lower value of $Q_{\mathrm{shallow}}$$=$1.6~MeV~nucleon$^{-1}$ (model 3; Figure~\ref{fig:cool} top right). However, the 2012 model calculations are less sensitive to this parameter because the first $\simeq$120 days of cooling were missed and it is this early cooling that most strongly constrains the required shallow heat \citep[e.g.,][]{brown08,degenaar2011_terzan5_3}. We can place $\simeq$2$\sigma$ upper limits of $Q_{\mathrm{shallow}}$$\simeq$2.2 and 2.9~MeV~nucleon$^{-1}$ for the 2012 and 2013 data, respectively.

The 2013 data allows us to fit for the maximum depth of shallow heating, for which we obtain $d_{\mathrm{shallow}}$$=$27.7. The 2012 data places almost no constraint on this parameter and we therefore chose to fix it to a value of $d_{\mathrm{shallow}}$$=$27.5 (model 3 and 4). We note that the exact value used for the 2012 data is unimportant and has very little effect on the inferred value of $Q_{\mathrm{shallow}}$ ($\pm0.1$~MeV~nucleon$^{-1}$ at most). This is again because there was no coverage in the first $\simeq$120 days of quiescence for this data set, which is the time period when the temperature is most sensitive to the shallow heating parameters. We can set a $\simeq$2$\sigma$ limit of $d_{\mathrm{shallow}}$$=$30.0 for the 2013 data.

As argued in Section~\ref{subsec:qspec}, it is possible that the temperature measured during the decay of an outburst already reflects the temperature of the crust, rather than that of the surface heated due to the infall of matter. We therefore also tried to describe the 2013 data with the decay data point included. We find that this is possible (model 2), i.e., the high temperature measured during the 2013 outburst decay can be achieved by crustal heating. We find that would put stronger requirements on the depth of the heating, yielding $d_{\mathrm{shallow}}$$=$27.2. Given that the shallow depth is defined in terms of $\log P$ and $y$$\simeq$$P/g$, this implies a factor of $\simeq$3 lower column depth than for the model run without the decay data point. The magnitude of the shallow heating is, however, hardly affected by including the decay point (cf. models 1 and 2 in Table~\ref{tab:model}). As can be seen in Figure~\ref{fig:cool} (top left), the effect of lowering $d_{\mathrm{shallow}}$ is to increase the start temperature of the cooling curve (blue dotted curve). When including the decay data point, we can set a $\simeq$2$\sigma$ upper limit of $d_{\mathrm{shallow}}$$=$27.4 for the 2013 data.

The XRT count rate light curves of the 2012 and 2013 data connect very smoothly when plotted as time since the peak of the outburst (Figure~\ref{fig:xrtlc20112013}), and the outburst energetics were remarkably similar (Table~\ref{tab:ob}). This suggests that it is worthwhile to model the two data sets together, i.e., regarding them as part of the same cooling curve so that we obtain a well-monitored base line of $>$1~yr after an accretion episode. We indeed find that a fit is possible (model 5--7; Figure~\ref{fig:cool} bottom left). The inferred parameters for the shallow depth are not very different from those obtained for the 2013 fits. This demonstrates that observations of the early cooling phase set the strongest constraints on the shallow heating. Nevertheless, comparing models 3 and 4 or 5 and 6, shows that the inferred shallow heat parameters are affected by the assumed core temperature. Therefore, measurements at later times are also needed to allow for the best possible crust cooling modelling.

It is possible that the core temperature of \source\ is lower than has been probed by observations. For this reason we also explored model calculations that allow the core temperature to fit the data. The 2013 observations alone are not useful for this purpose, because these only track the early cooling where the temperature is significantly elevated above the base level. We therefore modelled the 2012 and 2012+2013 data with the core temperature free. For the 2012 data set this a value of $T_{\mathrm{core}}$$=$$(1.6\pm0.1)\times10^8$ K (corresponding to $\mathrm{k}T^{\infty}$$\simeq$95 eV), i.e., lower than inferred from observations (model 4). This model run is shown as the black dashed curve in the top right panel of Figure~\ref{fig:cool}. Allowing for a lower core temperature improves the fit to the 2012 data (moving from ), although the level of improvement is not significant ($F$-test probability of 0.24). Combining the 2012 and 2013 data sets, we find a core temperature of $T_{\mathrm{core}}$$=$$(1.5\pm0.1)\times10^8$ K (corresponding to $\mathrm{k}T^{\infty}$$\simeq$92 eV), which provides a significant improvement over the higher core temperature inferred from observations (cf. model 5 and 6 in Table~\ref{tab:model}). 

To probe the effect of the uncertainty in the mass-accretion rate (Section~\ref{subsec:ob}), we also fit the 2012+2013 data with $\dot{M}$$=$$8\times10^{17}~\mdotg$, i.e., double our estimated value (model 7). This affects only the magnitude of the shallow heating, which is now half of that obtained for our basic model ($Q_{\mathrm{shallow}}$$=$1.2~MeV~nucleon$^{-1}$). This illustrates the degeneracy between the mass-accretion rate and the magnitude of the shallow heating (which is directly proportional to $\dot{M}$). Since the model run for the higher $\dot{M}$ completely overlays that of our basic run, we do not show it in Figure~\ref{fig:cool}. The depth inferred for the shallow heating is unchanged by assuming a different mass-accretion rate (cf. models 5 and 7). We note that the low $P_{\chi}$ value (Table~\ref{tab:model}) suggests that this is not a very good representation of the observational data, but we included it to demonstrate the high level of degeneracy between different parameters involved in the modelling.

\subsubsection{Numerical models for the 2015 data}
Having established that crust cooling is likely observed after the bright and long 2011 and 2013 outbursts, it is now interesting to study if crust cooling is also observable after the faint and short 2015 accretion episode. Unfortunately, there are not enough cooling points in 2015 to actually fit the data. However, we calculated predictions of what the cooling curve would look like for the outburst properties of 2015, and adopting the shallow heating parameters inferred for our basic 2012--13 calculation (model 5). 

Our prediction for the 2015 data (model 8) is shown in Figure~\ref{fig:cool} (bottom right) as the red curve. We find that this matches the observed temperatures quite well. Our model calculations show that after such a short and faint outburst, the temperature is lower and levels off much quicker ($\simeq$50~days) than for the longer and brighter outbursts ($\simeq$300~days). This is as we expect from the crust cooling paradigm (Section~\ref{sec:intro}), since the crust should be heated to a lesser extent during a short and faint outburst, hence it should be less hot and cool faster in quiescence.

Since there were also several decay observations close in time to the transition to quiescence, we calculated another cooling curve for the 2015 outburst (model 9), but now with the shallowed depth found for the 2013 data when including the decay (model 2). This model run, shown as the blue dotted curve in Figure~\ref{fig:cool} (bottom right), lies well below the decay data points. This suggests that even for a lower depth of the shallow heating, the 2015 outburst was not intense enough to result in the high temperatures inferred from the decay data points. If we try to fit the  temperature measurements inferred from the decay observations, the model completely overshoots the later data points. This likely indicates that, at least for the 2015 outburst, the temperature evolution during the decay was driven by accretion onto the stellar surface rather than crust cooling.


\vspace{-0.3cm}
\section{Discussion}\label{sec:discuss}
We investigated the hypothesis that the crust of the neutron star in \source\ is considerably heated during its accretion episodes and cools in quiescence. For that purpose we studied the decay and subsequent quiescent phase of the 2011, 2013 and 2015 outbursts, as observed with \swift/XRT. Daily monitoring with \swift/BAT and \maxi\ suggests that the 2011 and 2013 outbursts were strikingly similar in terms of brightness ($L_{\mathrm{2-50}}$$\simeq$$4\times10^{37}~\dist~\lum$) and duration ($t_{\mathrm{ob}}$$\simeq$8~weeks), whereas the 2015 one was both fainter ($L_{\mathrm{2-50}}$$\simeq$$7\times 10^{36}~\dist~\lum$) and shorter ($t_{\mathrm{ob}}$$\simeq$3~weeks). 

\swift\ covered the decay and first $\simeq$60~days of quiescence of the 2013 outburst. The transition into quiescence after the bright 2011 outburst was not captured by the XRT, but the source was densely monitored in 2012 covering $\simeq$120--375~days after accretion ceased. Strikingly, the 2012 and 2013 quiescent light curves line up very smoothly, indicting similar quiescent behaviour after two very similar outbursts. \swift\ also covered the decay of the 2015 outburst and followed it up to $\simeq$100~days in quiescence. Despite large differences in the outburst profile and energetics, the 2013 and 2015 outbursts decayed on a similar time scale of $\tau$$\simeq$1--2~days. However, notable differences are present in the count rate evolution of the subsequent quiescent phase. Immediately after the bright 2013 outburst the quiescent count rate is higher, and decays on a significantly longer time scale than after the fainter 2015 outburst.

\source\ is detected in quiescence with \swift\ at a luminosity of $L_{\mathrm{X}}$$\simeq$$10^{33}-10^{34}~\dist~\lum$ (0.5--10 keV). We modelled the XRT spectra as a combination of emission from the neutron star atmosphere and a power-law component, under the assumption that both the temperature and power-law normalisation varied between different epochs. This suggests that for each of the three outbursts studied in this work (2011, 2013, and 2015) there is an overall decrease in neutron star temperature with time progressing after accretion ceased. After the faint and short 2015 outburst the measured temperatures were lower, and decreased more rapidly, than after the long and bright 2011 and 2013 outbursts.

\subsection{Crust cooling and shallow heating in \source}\label{subsec:aqlcool}
The temperature evolution of \source\ after different outbursts is consistent with the crust cooling paradigm. Since a longer and brighter outburst (such as observed in 2011 and 2013) should result in more intense heating, a higher temperature and longer decay time scale should be observed than a shorter and fainter outburst (such as in 2015). We therefore self-consistently modelled the observational data with a neutron star thermal evolution code. Using outburst constraints from \swift/BAT and \maxi\ monitoring, and assuming standard microphysics input, we can indeed successfully describe the temperature evolution observed for \source\ in quiescence. This supports the hypothesis that the crust of the neutron star is significantly heated during accretion outbursts, and can be observed to cool in subsequent quiescent phases.

From our theoretical modelling we can place some constraints on the shallow heat release in the crust. We find that a shallow heat source of $Q_{\mathrm{shallow}}$$\simeq$$1.2-3.2$ MeV~nucleon$^{-1}$ matches the data of \source, depending on the assumed mass-accretion rate and core temperature. We can set a $\simeq$2$\sigma$ upper limit of $Q_{\mathrm{shallow}}$$\lesssim$$3.4$ MeV~nucleon$^{-1}$. This is of the same magnitude as the extra heat energy needed to model the crust cooling curves of \ks, \mxb, \exo, and \igr\ \citep[][]{brown08,degenaar2013_ter5,degenaar2014_exo3}, albeit lower than the $Q_{\mathrm{shallow}}$$\simeq$10 MeV~nucleon$^{-1}$ required for the exceptionally hot neutron star in \maxisource\ \citep[][]{deibel2015}. 

Dense coverage around the transition from outburst to quiescence in 2013 also allows us to place some constraint on the depth of the shallow heat release. From our modelling to the combined 2012--2013 data set we find a maximum depth for the shallow heat of $\log P$$=$27.5 ($P$$\simeq$$3\times10^{27}~\mathrm{erg~cm}^{-3}$), which corresponds to a column depth of $y$$\simeq$$10^{13}~\mathrm{g~cm}^{-2}$. This is broadly consistent with expectations from superbursts, which are thought to occur at $y$$\simeq$$10^{12}~\mathrm{g~cm}^{-2}$ and require additional heating to get the temperature sufficiently high to achieve ignition at this depth \citep[e.g.,][]{cumming06}. We note that \source\ is itself not a superburster.

The standard nuclear heating of neutron star crusts is proportional to the mass-accretion rate and therefore scales directly with the outburst properties \citep[e.g.,][]{haensel2008}. However, the extra shallow heating inferred for several neutron stars may not necessarily be proportional to the mass-accretion rate and could possibly vary for different outbursts. For instance, \citet{deibel2015} found that in the crust-cooling source \maxisource\ the exceptionally strong shallow heating found during the main outburst could not have been operating at similar strength during a bright re-flare (with a duration of $\simeq$60 days and an average luminosity of $L_{\mathrm{X}}$$\simeq$$4\times10^{37}~\lum$) that occurred $\simeq$170 days into quiescence \citep[][]{homan2014}. Furthermore, a rapid cessation of X-ray bursts (likely indicating sudden intense heating of the outer crustal layers) in 4U 1820--30 after a spectral state transition, suggests that shallow heating may depend on the accretion geometry \citep[][]{zand2012}. 

A distinction in shallow heating for different outbursts is not apparent from our analysis of \source. Comparing its 2011 and 2013 outbursts with that of 2015 suggests that the former, brighter accretion episodes were spectrally much softer (i.e., the flux measured by \maxi\ was much higher in 2011 and 2013 than in 2015, whereas the \swift/BAT fluxes were similar; Figure~\ref{fig:maxibatlc201120132015}). However, we find that the quiescent behaviour can be explained by the same shallow heating parameters for the two different types of outburst (Table~\ref{tab:ob}). Nevertheless, only limited data points were available for the quiescent state following the 2015 outburst, so  we cannot draw firm conclusions about whether or not there is a difference in shallow heating after different outbursts.

\subsection{Comparison with the 2010 outburst}\label{subsec:2010}
It is worthwhile to compare our results with reports of the decay of the 2010 outburst, which was monitored closely with \chan\ and \xmm\ \citep[][]{campana2014}. From \maxi\ and \swift/BAT coverage of the outburst we estimate an average luminosity of $L_{2-50}$$\simeq$$2\times10^{37}~\lum$, and a duration of $\simeq$30 days, suggesting that it falls in between the bright 2011/2013 and the faint 2015 outbursts in terms of energetics (cf. Table~\ref{tab:ob}). The neutron star temperatures reported for 2010 range from $kT^{\infty}$$\simeq$$150-195$~eV during the decay to $kT^{\infty}$$\simeq$$115-130$~eV during the first $\simeq$30~days of quiescence. Unfortunately monitoring stopped there. 

The temperature measured during the last observation reported in \citet{campana2014} was $kT^{\infty}$$\simeq$120~eV, whereas the lowest temperature we detect with \swift\ is $kT^{\infty}$$\simeq$104~eV if we assume the same mass and radius as in that study ($M$$=$$1.4~\Msun$ and $R$$=$10~km). For comparison, $\simeq$30~days post-outburst we measure a temperature of $kT^{\infty}$$\simeq$140~eV for the brighter 2013 outburst and $kT^{\infty}$$\simeq$$110$~eV for the fainter 2015 outburst. The results for the 2010 outburst would thus fall right in between, as would be expected within the crust cooling paradigm based on the relative energetics of the 2010, 2013 and 2015 outbursts.

Although the quiescent behaviour after different outbursts is thus consistent with the expectations in the framework of crust cooling, it is of note that despite the very different outburst profiles and energetics, the temperatures inferred during the decay of the 2010, 2013, and 2015 outbursts are very similar: $kT^{\infty}$$\simeq$$150-200$~eV. This may argue in favour of the thermal emission seen during the decay of the outbursts being dominated by surface heating due to the infall of matter, rather than cooling of the crust. Our modelling of the 2015 data also pointed in this direction: The temperatures observed during the decay were much higher than expected for the crust for the 2015 outburst energetics.

\subsection{Comparison with other crust-cooling sources}\label{subsec:compare}
\source\ is the third neutron star LMXB with an outburst of $<$1~yr that shows evidence for crust cooling, after \igr\ and \terdrie\ \citep[which are both located in the globular cluster Terzan 5;][]{degenaar2011_terzan5_2,degenaar2011_terzan5_3,degenaar2013_ter5,degenaar2015_ter5x3}. \igr\ (an 11-Hz X-ray pulsar) accreted for $t_{\mathrm{ob}}$$\simeq$$0.17$~yr at $\dot{M}$$\simeq$$2\times 10^{17}~\mdotg$ when it returned to quiescence in 2010, whereas \terdrie\ was in outburst for $t_{\mathrm{ob}}$$\simeq$$0.15$~yr in 2012 with $\dot{M}$$\simeq$$1\times 10^{17}~\mdotg$. Their pre-outburst temperatures were $kT^{\infty}$$\simeq$$74$ and $\simeq$$90$~eV, respectively. 

Long-term monitoring has shown that the neutron star crust in \igr\ continues to cool $>$5~yr after its outburst, whereas \terdrie\ appeared to have completely cooled within $\simeq$100~days after entering quiescence \citep[][]{degenaar2015_ter5x3}. \igr\ requires a shallow heat source of order $Q_{\mathrm{shallow}}$$\simeq$$1$~MeV~nucleon$^{-1}$ \citep[][]{degenaar2011_terzan5_2,degenaar2013_ter5}. For \terdrie\ no shallow heating is needed, although as much as $Q_{\mathrm{shallow}}$$\simeq$$1.4$~MeV~nucleon$^{-1}$ was allowed by the data \citep[][]{degenaar2015_ter5x3}. The cooling time scale of \source\ after its 2011/2013 outbursts lies in between the values of these two sources, and its shallow heating seems to be of similar magnitude or slighter higher (depending on the assumed mass-accretion rate and core temperature).

The 2011 and 2013 outbursts of \source\ were of similar length as those of \igr\ and \terdrie, whereas its mass-accretion rate was a factor $\simeq$2--4 higher than for the other two. However, the quiescent base level of \source\ is also considerably higher. With a higher core temperature, a less steep temperature gradient develops for a given mass-accretion rate, leading to faster cooling. This may be the reason that despite its lower mass-accretion rate the cooling timescale in \igr\ is much longer than in \source\ ($>$5~yr vs $\simeq$1~yr), as its base temperature is much lower ($kT^{\infty}$$\simeq$$74$~eV vs $kT^{\infty}$$\simeq$$100$~eV). Whereas \source\ also has a higher base temperature than \terdrie\ ($kT^{\infty}$$\simeq$$90$~eV), its higher mass-accretion rate and possibly stronger shallow heating may be the reason that \source\ still has a longer cooling time scale ($\simeq$1~yr vs $\simeq$100~days).

\subsection{On the possibility of residual accretion}\label{subsec:resacc}
We cannot exclude that long-term decay trends seen in the quiescent thermal emission of \source\ are caused by low-level accretion, since there are no ways to conclusively determine this. However, it is not obvious why such residual accretion would be more intense after a bright outburst than after a fainter one, as suggested by our \swift/XRT analysis. On the other hand, the observed behaviour is consistent with expectations if cooling of the accretion-heated crust is dominating the quiescent light curve evolution.

The fact that the evolution of the thermal emission of \source\ can be successfully modelled as crust cooling does not necessarily imply, however, that residual accretion does not occur. In fact, the two flares seen in 2012 provide strong evidence that there is still matter reaching the neutron star surface, either sporadically or continuously \citep[][]{cotizelati2014}. Nevertheless, it is possible that the heat flow from the interior of the neutron star maintains the surface at a higher temperature than any possible continuous low-level accretion \citep[see also][]{walsh2015}. If so, thermal emission from the accretion-heated crust may drive the overall evolution in the quiescent emission, even if residual accretion occurs. Sporadic, more intense spurts of accretion may then appear as variations on top of the underlying cooling. This has also been proposed for \xte\ and \maxisource, where accretion flares seem to leave the underlying decay from crust cooling unaffected, as seen for \source\ \citep[][]{fridriksson2011,homan2014}.

It is worth noting that the 2012 quiescent observations of \source\ (with the exception of the last set of 7 observations) had a significant power-law contribution, whereas the 2013 and 2015 quiescent data were completely thermally dominated. It is striking then, that two accretion flares were seen in 2012, but not in the other years (although coverage was more extended in 2012 than in 2013 and 2015). This could indicate that there was continuous low-level accretion in 2012, but not in 2013 and 2015. Nevertheless, the neutron star temperature evolves along the same trend in 2012  and 2013. This suggests that the temperature of the crust may have indeed been higher than that generated by accretion onto the surface, so that the observed temperature is dominated by the thermal state of the crust rather than that of the accretion-heated surface. 

Our crust cooling modelling suggest that the true base temperature may be lower ($kT^{\infty}$$\simeq$92~eV) than the lowest value ever reported for \source\ ($kT^{\infty}$$\simeq$100~eV for the same $M$ and $R$). This could imply that the lowest observed level of $kT^{\infty}$$\simeq$100~eV is powered by low-level accretion, or that the crust does not have time to cool completely before a new outburst starts. Our simulations suggest that it takes several years for the neutron star to cool to $kT^{\infty}$$\simeq$92~eV, which is much shorter than the recurrence time of outbursts in \source\ \citep[e.g.,][]{campana2013}. Moreover, during our lowest temperature measurement of $kT^{\infty}$$\simeq$100~eV, the quiescent X-ray spectrum was completely thermally dominated (i.e., there was no need to include a power-law spectral component in the fits). This may suggest that residual accretion is absent, hence that this level reflects the temperature of the crust rather than that of the accretion-heated surface.

\subsection{Future prospects for crust cooling studies}\label{subsec:future}
\source\ is in outburst very often, the neutron star spin frequency is known \citep[$\nu_{\mathrm{s}} \simeq 550$~Hz;][]{casella2008}, and a distance has been inferred from X-ray burst studies \citep[$D\simeq$5~kpc;][]{rutledge2001}. Therefore, the source may serve as a promising asset to advance our understanding of shallow heating in neutron star crusts, as it provides the opportunity to measure crust cooling curves after different outbursts. Since the distance, mass, radius and spin period do not change, these cannot cause any observed differences in cooling behaviour, as is the case when comparing different sources among each other. This could be a powerful tool to break current degeneracies in crustal heating and cooling models, and to gain more insight into the crust microphysics.

Our results on \source\ show that crust cooling can be observable in spite of a relatively high base temperature. 4U 1608--52 is another neutron star LMXB that is in outburst roughly once every $\simeq$1--2~years \citep[e.g.,][]{simon2004}, has a distance estimate \citep[$D$$\simeq$3.6~kpc; e.g.,][]{poutanen2014_4u1608} and a neutron star spin measurement \citep[$\nu_{\mathrm{s}}$$\simeq$$620$~Hz;][]{muno2001}. Moreover, this source also displayed a superburst, which gives additional constraints on the shallow crustal heating \citep[][]{keek2008_1608}. 4U 1608--52 could therefore also serve as a potential target for future crust cooling studies, although a disadvantage is that it often lingers at a low luminosity level after an outburst before decaying into quiescence \citep[e.g.,][]{simon2004}. During that time much of the crustal heat may already be lost \citep[e.g.,][]{deibel2015}, complicating both observations and theoretical modelling. 

Modelling the thermal evolution of \source\ has clearly demonstrated that dense sampling of the decay of an outburst, and first tens of days in quiescence, places the strongest constraints on the depth and magnitude of the elusive shallow crustal heating. Nevertheless, the shallow heating parameters inferred from the modelling are somewhat sensitive to the core temperature, which therefore needs to be measured as well. This can be achieved at later times, $\gtrsim$1~yr post-outburst, and preferably via multiple observations to ensure that the neutron star has fully cooled, and to avoid catching it at an elevated temperature due to an accretion flare.

\vspace{-0.3cm}
\section*{Acknowledgements}
ACW was funded by an undergraduate bursary awarded to ND by the Royal Astronomical Society. ND acknowledges support via an EU Marie Curie Intra-European fellowship under contract no. FP-PEOPLE-2013-IEF-627148. DA acknowledges support from the Royal Society. ML was supported by the Spanish Ministry of Economy and Competitiveness under the grant AYA2013-42627. This work made use of data supplied by the UK \swift\ Science Data Centre at the University of Leicester. We also made use of MAXI data provided by RIKEN, JAXA and the MAXI team, and public light curves from the \swift/BAT transient project. The authors are grateful to the anonymous referee for valuable comments that helped improve this manuscript.

\vspace{-0.3cm}
\footnotesize{
\bibliographystyle{mn2e}

}

\end{document}